\documentclass[aps,prx,article,twocolumn,preprintnumbers,amsmath,amssymb,superscriptaddress]{revtex4-2}
\usepackage[english]{babel}
\usepackage[T1]{fontenc}
\usepackage{braket}
\usepackage{float}

\newcommand\ov{\over}

\newcommand{\be}{\begin{equation}}
\newcommand{\ee}{\end{equation}}
\newcommand{\bea}{\begin{eqnarray}}
\newcommand{\eea}{\end{eqnarray}}
\newcommand{\bega}{\begin{gather}}

\newcommand\vev[1]{{\ensuremath{\left\langle{#1}\right\rangle}}}

\newcommand\sD{{\ensuremath{{\mathcal D}}}}

\newcommand\sH{{\ensuremath{{\mathcal H}}}}

\newcommand\sM{{\ensuremath{{\mathcal M}}}}

\usepackage{amsfonts}
\usepackage{graphicx}
\usepackage{subfigure}	
\usepackage{bbold}
\usepackage{simpler-wick}
\usepackage{dcolumn}
\usepackage{bm}
\usepackage[colorinlistoftodos]{todonotes}
\usepackage[colorlinks=true, allcolors=blue]{hyperref}
\usepackage[titletoc]{appendix}
\usepackage{epsfig}
\usepackage{float}
\usepackage{comment}

\newcommand{\vb}[1]{\mathbf{#1}}

\newcommand{\kbf}      {\textbf{k}}

\newcommand{\qbf}      {\textbf{q}}

\newcommand{\rbf}      {\textbf{r}}

\newcommand{\eps} {{\mathbf{\varepsilon}}}

\makeatletter
\newsavebox{\@brx}
\newcommand{\llangle}[1][]{\savebox{\@brx}{\(\m@th{#1\langle}\)}%
  \mathopen{\copy\@brx\mkern2mu\kern-0.9\wd\@brx\usebox{\@brx}}}
\newcommand{\rrangle}[1][]{\savebox{\@brx}{\(\m@th{#1\rangle}\)}%
  \mathclose{\copy\@brx\mkern2mu\kern-0.9\wd\@brx\usebox{\@brx}}}
\makeatother

\allowdisplaybreaks
\date{\today}
\begin{abstract}
Characterizing quantum materials is essential for understanding their microscopic interactions and advancing quantum technology. X-ray photon correlation spectroscopy (XPCS) with coherent X-ray sources offers access to higher-order correlations, but its theoretical basis, the Siegert relation, is derived from dynamical light scattering with independent classical scatterers, and its validity for XPCS remains unexamined. Here we present a microscopic quantum theory of XPCS derived from elecron-photon interaction Hamiltonians, introducing four configurations tied to distinct fourth-order electron-density correlation functions. We examine the validity of the Siegert relation and derive a generalized Siegert relation. Notably, the Siegert relation breaks down even in non-interacting Fermi gas due to exchange interactions. Furthermore, density matrix renormalization group calculations on 1D Kitaev chain reveal oscillatary signatures that can distinguish topologically trivial phases from topological phases with Majorana zero modes. Our work provides a robust theoretical foundation for XPCS and highlights the value of higher-order correlations in advanced X-ray and neutron sources for probing quantum materials.

\end{abstract}

\begin{document}

\title{Quantum Theory of X-ray Photon Correlation Spectroscopy}

\affiliation{%
 Quantum Measurement Group, MIT, Cambridge, MA 02139, USA}
\affiliation{Department of Physics, MIT, Cambridge, MA 02139, USA}
\affiliation{Department of Nuclear Science and Engineering, MIT, Cambridge, MA 02139, USA}
 \affiliation{Department of Chemistry, MIT, Cambridge, MA 02139, USA}
\affiliation{Department of Chemistry, Emory University, Atlanta, GA 30322, USA}
\affiliation{These authors contributed equally to this work.}

\author{Phum Siriviboon}
\thanks{Corresponding authors. \href{mailto:psirivib@mit.edu}{psirivib@mit.edu}, \href{mailto:mingda@mit.edu}{mingda@mit.edu}}
\affiliation{%
 Quantum Measurement Group, MIT, Cambridge, MA 02139, USA}
\affiliation{Department of Physics, MIT, Cambridge, MA 02139, USA}
\affiliation{These authors contributed equally to this work.}

\author{Chu-Liang Fu}
\affiliation{%
 Quantum Measurement Group, MIT, Cambridge, MA 02139, USA}
\affiliation{Department of Nuclear Science and Engineering, MIT, Cambridge, MA 02139, USA}
\affiliation{These authors contributed equally to this work.}

\author{Michael Landry}
\affiliation{%
 Quantum Measurement Group, MIT, Cambridge, MA 02139, USA}
\affiliation{Department of Physics, MIT, Cambridge, MA 02139, USA}
\affiliation{These authors contributed equally to this work.}

\author{Ryotaro Okabe}
\affiliation{%
 Quantum Measurement Group, MIT, Cambridge, MA 02139, USA}
 \affiliation{Department of Chemistry, MIT, Cambridge, MA 02139, USA}

\author{Denisse Córdova Carrizales}
\affiliation{%
 Quantum Measurement Group, MIT, Cambridge, MA 02139, USA}
 \affiliation{Department of Nuclear Science and Engineering, MIT, Cambridge, MA 02139, USA}

\author{Yao Wang}
\affiliation{Department of Chemistry, Emory University, Atlanta, GA 30322, USA}

\author{Mingda Li}
\thanks{Corresponding authors. \href{mailto:psirivib@mit.edu}{psirivib@mit.edu}, \href{mailto:mingda@mit.edu}{mingda@mit.edu}}
\affiliation{%
 Quantum Measurement Group, MIT, Cambridge, MA 02139, USA}
\affiliation{Department of Nuclear Science and Engineering, MIT, Cambridge, MA 02139, USA}
\maketitle

\section{Introduction}

The characterization of quantum materials plays a critical role in understanding their microscopic interaction mechanisms. Recent decades have witnessed an explosion of diverse quantum material families, emerging from the intricate interplay of symmetry, topology, and electron correlation \cite{rise2016, keimer2017physics, giustino20212021}. These materials span a wide range of physics, including low-dimensional systems such as graphene and Moiré structures \cite{geim2009graphene,cao2018correlated}; correlated phases such as unconventional superconductors \cite{norman2011challenge,stewart2017unconventional}, strange metals \cite{phillips2022stranger}, and Kondo insulators \cite{checkelsky2024flat}; topological phases such as topological insulators and Weyl semimetals \cite{hasan2010colloquium,qi2012topological, RevModPhys.90.015001,manna2018heusler}, topological superconductors \cite{RevModPhys.89.041004,sato2017topological,mandal2023topological}, and quantum spin liquids \cite{savary2016quantum,RevModPhys.89.025003,broholm2020quantum,clark2021quantum}. Furthermore, recent advances in machine learning have allowed the generation of millions of new functional and quantum materials \cite{merchant2023scaling,okabe2024structural}. 

However, despite the rapid growth in the discovery of new quantum materials, the techniques available to characterize them remain limited. Many theoretical phases have not been concretely established experimentally due to a lack of clear experimental signatures. For instance, despite extensive efforts to identify topological qubits of Majorana zero modes in topological superconductors using high-resolution scanning tunneling spectroscopy (STS), collecting conclusive evidence remains challenging due to contamination by spurious signals \cite{PhysRevResearch.2.013377,cheng2024machine}. It is similarly difficult to identify quantum spin liquids despite intensive searches over the past decade using techniques like thermal transport and neutron scattering. Only a few dozen quantum spin liquid candidates have been identified \cite{broholm2020quantum,clark2021quantum}, and clear identification remains difficult \cite{PhysRevX.9.041051} because some results are prone to disorder effects \cite{PhysRevMaterials.8.014402,PhysRevB.108.144419}. Another bottleneck to materials characterization is that many powerful characterization techniques are restricted to probing lower-order correlation functions, leaving higher-order correlations largely unexplored. For instance, angle-resolved photoemission spectroscopy (ARPES) and STS are extremely powerful but only access single-particle spectral functions. 
Magnetic neutron scattering and optical conductivity measure second-order spin-spin correlation and current-current correlation, respectively. These limitations highlight the pressing need for advanced characterization techniques that can probe higher-order correlation functions.

X-ray photon correlation spectroscopy (XPCS) is a powerful technique uniquely capable of measuring higher-order correlations of materials. Over the past two decades, the rapid advancements in synchrotron-based coherent X-ray technology have propelled XPCS into prominence, particularly for studying mesoscopic or nanoscale dynamics in systems such as solid solutions \cite{stana2013studies}, soft matter \cite{lu2010temperature}, metallic glasses \cite{das2019stress}, charge density waves \cite{campi2022nanoscale}, and magnetic domains \cite{ricci2020intermittent}. The study of these dynamics is achieved by capturing the time-resolved fluctuations in speckle patterns produced by coherent X-ray beams. However, XPCS has primarily been used for classical charge systems. Its theoretical foundation is based on the so-called Siegert relation, which is the foundation for dynamic light scattering (DLS)---also known as photon correlation spectroscopy (PCS). PCS assumes that the scattering comes from a large number of independently and randomly distributed scatterers, resulting in the electric field being modeled as a Gaussian stochastic variable. However, X-rays directly interact with electrons, which can result in an instantaneous breakdown of the Siegert relation when the electrons are not independent. To the best of our knowledge, the applicability of the Siegert relation in the X-ray regime has not been examined. Since XPCS holds significant potential for probing higher-order correlations in quantum materials, which is evident from the fact that the first beamline in the fourth-generation synchrotron X-ray Advanced Photon Source (APS-U) is dedicated to XPCS, it is essential to develop a microscopic theoretical framework for XPCS, which can offer a solid foundation of XPCS for future synchrotron and free-electron lasers.

In this work, we present a comprehensive microscopic quantum theory of XPCS. Instead of making \textit{ad hoc} hypotheses, we derive the XPCS correlation functions directly from the photon-electron interaction Hamiltonians using a minimal-coupling scheme, which is valid under the weak field approximations. To ensure the theory’s applicability to various X-ray sources, including continuous-beam synchrotrons and free-electron lasers with pulse trains, we formulate the X-ray probe structure in a general manner using second quantization and coherent state language. XPCS’s momentum-dependent scattering and insights from Hanbury Brown and Twiss (HBT) experiments \cite{brown1956correlation} in quantum optics lead us to propose four distinct XPCS configurations (one of them being the known configuration), each corresponding to a unique fourth-order correlation function. Under the plane-wave approximation, these correlation functions are reduced to fourth-order electron density correlation functions. We examine the conditions where the Siegert relation breaks down, showing that it can fail under two scenarios: in the case of classical scatterers with special momentum constraint (``classical breakdown'') and in electron liquids due to exchange correlations (``quantum breakdown''). We derive a generalized Siegert relation which incorporates two-momentum and compute the density correlation functions in non-interacting Fermi gas. An oscillatory signature of XPCS arises due to the electron exchange, indicating the power of XPCS to study exchange-correlation effects. Finally, we compute the correlation functions for a topological superconductor, using the 1D Kitaev chain as toy model, and show that XPCS can effectively distinguish between topologically trivial and topological phases.

The organization of this paper is as follows: We introduce XPCS setup with a comprehensive theoretical foundation in Sec.\,\ref{sec:setup}. In particular, the four XPCS configurations are introduced in Sec.\,\ref{sec:setup}D. Next, we dedicate Sec.\,\ref{sec:siegert} to discuss the breakdown and the generalization of the Siegert relation. In Sec.\,\ref{sec:qes}, we compute the electron correlation functions in the non-interacting Fermi gas and 1D Kitaev chain.  We conclude our paper with a brief outlook in Sec.\,\ref{sec:con}.

\section{Setup of XPCS measurement}\label{sec:setup}
\subsection{System Hamiltonian Setup}

We consider a system that consists of electron and photon degrees of freedom. The total Hamiltonian of the composite system is given by 
\begin{eqnarray}\label{eq:totalHam}
    \mathcal{H}= \mathcal{H}_0 + \mathcal{H}_{\rm int}  = \mathcal{H}_{\rm e} + \mathcal{H}_{\rm ph} + \mathcal{H}_{\rm int},
\end{eqnarray}
where $\mathcal{H}_{\rm e}$ is the electronic Hamiltonian, $\mathcal{H}_{\rm ph} =\sum_{\mathbf{p}\bm{\eps}} \hbar \omega_{\mathbf{p}} a^\dagger_{\mathbf{p}\bm{\eps}} a_{\mathbf{p}\bm{\eps}}$ is the free photon Hamiltonian, and $\sH_{\rm int}$ is the electron-photon interaction Hamiltonian. The operator $a_{\mathbf{p}\bm{\eps}}$ (or $a^\dagger_{\mathbf{p}\bm{\eps}}$) annihilates (or creates) a photon state with wavevector $\mathbf{p}$, energy $\hbar \omega_{\mathbf{p}}$ and polarization $\bm{\eps}$, which satisfies the canonical commutation relation $[a_{\mathbf{p}\bm{\eps}},a^\dagger_{\mathbf{p}'\bm{\eps}'}]=\delta_{\mathbf{p}\mathbf{p}'}\delta_{\bm{\eps}\bm{\eps}'}$.  Additionally, $\mathcal{H}_0=\mathcal{H}_{\rm e} + \mathcal{H}_{\rm ph}$  denotes the total electron-photon Hamiltonian without considering their interactions, which is defined for later convenience. Under the scheme of minimal coupling and quantized vector potential (Eq. \eqref{eq:Art}), the electron-photon interaction Hamiltonian $\mathcal{H}_{\rm int}$ can be written in a second quantized form as
\begin{eqnarray}\label{eq:Hamint}
     \mathcal{H}_{\rm int}\!\!\!&=&\!\!\sum_{\mathbf{p}\bm{\eps}}\sum_{\kbf\alpha\beta} \left[ D^{\alpha\beta\bm{\eps}}_{\kbf,\mathbf{p}} c^{\dagger}_{\kbf+\mathbf{p}\beta}c_{\kbf\alpha}a_{\mathbf{p}\bm{\eps}} +h.c.\right] \nonumber\\
    && \!\!\!\!\!\!+\sum_{\kbf\alpha\beta}\!\sum_{\mathbf{q}\mathbf{p}_{i}\atop\bm{\eps}_{i}\bm{\eps}_f} \!\left[M^{\alpha\beta\bm{\eps}_{i}\bm{\eps}_f}_{\kbf,\mathbf{p}_{i},\mathbf{q}}c^{\dagger}_{\kbf+\mathbf{q}\beta}c_{\kbf\alpha}a_{ \mathbf{p}_{i} - \mathbf{q} \bm{\eps}_{f}}^\dagger a_{\mathbf{p}_{i}\bm{\eps}_{i}} \!  \right]\,\nonumber\\ &&\!\!\!\!\!\!+\, \textrm{two-photon absorption/emission terms},
\end{eqnarray}
where the first line with the linear dependence on $a_{\mathbf{p}\bm{\eps}}$ (or $a^{\dagger}_{\mathbf{p}\bm{\eps}}$) represents the single photon absorption (or emission) process, while the second line with quadratic dependence on photon creation and annihilation operators denotes the photon scattering process. Here, the operator $c_{\kbf}$ (or $c^{\dagger}_{\kbf}$) annihilates (or creates) an electron state with wavevector $\kbf$; $\alpha,\beta$ are the internal degrees of freedom of electrons, e.g., band index and spin, and the summed photon momenta labels ($\mathbf{p}$, $\mathbf{p}_i$, $\mathbf{q}$) represent the different scattering events in Fock space of the total Hilbert space. Before the light-matter interaction, we assume that the electronic state is in thermal equilibrium while the photon state is coherent (to be discussed in Sec.\,\ref{sec:setup}\,B).
The absorption matrix element $D^{\alpha\beta\bm{\eps}}_{\kbf,\mathbf{p}}$ can be written in single-particle basis as  
\begin{eqnarray}\label{eq:dipoleMatrixElement}
    D^{\alpha\beta\bm{\eps}}_{\kbf,\mathbf{p}}   = -\frac{ie \hbar^{3/2}}{m\sqrt{2\bm{\eps}_0\omega_{\mathbf{p}} V}} \int 
 d\rbf e^{i{\bf p\cdot r}} \psi^*_{\kbf+\mathbf{p}\beta}(\rbf)  \bm{\eps}\cdot\nabla \psi_{\kbf\alpha}(\rbf),
\end{eqnarray}
where $V$ is the system volume, $m$ is electron mass, $\bm{\eps}_0$ is the dielectric constant of vacuum, and $\psi_{\kbf\alpha}$ are single-particle electron orbitals. The scattering matrix element $M^{\alpha\beta\bm{\eps}_{i}\bm{\eps}_f}_{\kbf,\mathbf{p}_{i},\mathbf{q}}$ is given by
\begin{eqnarray}\label{eq:scatteringMatrixElement}
M^{\alpha\beta\bm{\eps}_{i}\bm{\eps}_f}_{\kbf,\mathbf{p}_{i},\mathbf{q}}
\!&=&\!\frac{e^2\hbar}{2m\bm{\eps}_0 V\sqrt{\omega_{\mathbf{p}_{i}}\omega_{\mathbf{p}_{i} - \mathbf{q}}}}\bm{\eps}_f^* \cdot \bm{\eps}_{i}\nonumber\\
    &\times&
    \int d\rbf \psi^*_{\kbf+\qbf\beta}({\rbf})\psi_{\kbf\alpha}(\rbf)
    e^{i\mathbf{q}\cdot{\rbf}}\,,
\end{eqnarray}
which reflects the non-resonant scattering from the electron charge multipoles measured by the incident X-ray beam. For non-resonant X-ray scattering process, the contribution from the photon absorption and emission term in $\mathcal{H}_{\rm int}$ vanishes. As a result, in this work, we only focus on the scattering contribution in $\mathcal{H}_{\rm int}$ and neglect the absorption or emission effect, 
\begin{eqnarray}\label{eq:Hamint}
    \mathcal{H}_{\rm int} \approx \sum_{\kbf\alpha\beta}\!\sum_{\mathbf{p}_{i}\mathbf{q}\atop\bm{\eps}_{i}\bm{\eps}_f} \!\left[M^{\alpha\beta\bm{\eps}_{i}\bm{\eps}_f}_{\kbf,\mathbf{p}_{i},\mathbf{q}}c^{\dagger}_{\kbf+\mathbf{q}\beta}c_{\kbf\alpha}a_{ \mathbf{p}_{i} - \mathbf{q} \bm{\eps}_{f}}^\dagger a_{\mathbf{p}_{i}\bm{\eps}_{i}} \right].
\end{eqnarray}

\subsection{General Theory of Photon Probe of XPCS}
\label{sec:photon}
To develop a full microscopic quantum theory for XPCS, in addition to the Hamiltonian given by Eqs.\,\eqref{eq:totalHam} and \eqref{eq:Hamint}, we also must identify the photon states that can represent general coherent X-ray beams. Throughout this work, we employ $\ket{\psi}_{\rm e}$ to describe the electron states of the sample, $\ket{\phi}_{\rm ph}$ to represent the photon states of the X-ray beam, and $|\Psi\rangle$ to represent the total state of the whole system. Before any interaction has taken place, we assume that the electron and photon exist in a separable pure state, which means that the initial state of the full system has $|\Psi_0\rangle = |\psi_0\rangle _{\rm e} \otimes |\phi_0\rangle _{\rm ph}$, where $|\psi_0\rangle _{\rm e}$ and $|\phi_0\rangle _{\rm ph}$ are the initial electron state and the initial photon state, respectively. We can also generalize the initial pure state $\ket{\Psi_0}$ to a mixed state $\rho = \frac{e^{- \beta \mathcal{H}_{\rm e}}}{Z} \otimes |{\phi}\rangle_{\rm ph}\langle{\phi}|_{\rm ph}$. $Z$ is the corresponding partition function as the normalization factor.

We define the initial photon state through the time profile of the photon beams. There are two main types of photon beams deployed as coherent X-ray sources: (1) the photon real-time amplitude is a continuous constant mode, as in the case of a synchrotron source; or (2) the photon forms a pulse train, as in the case of a free-electron laser. Therefore, we provide a self-contained construction for defining the initial generic coherent photon state~$\ket{\phi}_{\rm ph}$. 

We start from a vector potential operator for the photon,
\begin{equation}\label{eq:Art}
    \mathbf{A}(\rbf,t) = \sum_{\mathbf{p}\bm{\eps}}  \sqrt{\frac{\hbar}{2\bm{\eps}_0\omega_\mathbf{p} V}} \left[ e^{i{\mathbf{p}\cdot \mathbf{r}}} a_{\mathbf{p}\bm{\eps}} (t) \bm{\eps} + h.c. \right],
\end{equation}
which is defined in the Heisenberg picture. For simplicity and later convenience, the summation over the two polarization states is denoted by the polarization vector $\bm{\eps}$ itself, instead of using the more common notation $\bm{\eps}_{\lambda}$ with $\lambda=1,2$. For free photons, the time-dependent annihilation operator in the Heisenberg picture can be written in terms of the time-independent operator $a_{\vb{p}\bm{\eps}}(t) = e^{- i\omega_{\mathbf{p}}t} a_{\vb{p}\bm{\eps}}$. 

To model the quantum state of coherent X-ray beams with a specific real-time amplitude, we adopt the coherent state formalism. A generic photon coherent state can be written as:
\begin{align}
        \label{eq:coherent}
    \ket{\phi}_{\rm ph} &= \frac{1}{\mathcal{N}^{1/2}}\exp \Bigl( {\sum_{\mathbf{p}\bm{\eps}}s_{\mathbf{p}\bm{\eps}}a^\dagger_{\mathbf{p}\bm{\eps}}} \Bigr) \ket{0}_{\rm ph} \notag \\
    &= \exp \Bigl( \sum_{\mathbf{p}\bm{\eps}} \bigl(s_{\mathbf{p}\bm{\eps}}a^\dagger_{\mathbf{p}\bm{\eps}}-s^{*}_{\mathbf{p}\bm{\eps}}a_{\mathbf{p}\bm{\eps}}\bigr) \Bigr) \ket{0}_{\rm ph},
\end{align}

\noindent in which $\ket{0}_{\rm ph}$ is the photon vacuum, $\mathcal{N}^{1/2} = {\exp \left ({\frac{1}{2} \sum_{\mathbf{p}\bm{\eps}} |s_{\mathbf{p}\bm{\eps}}|^2} \right )}\ $ is the normalization factor,  $s_{\mathbf{p}\bm{\eps}}$ is the amplitude which is determined by the temporal and spatial structure of the photon beam; the second line of Eq.\,\eqref{eq:coherent} is the displacement operator representation.   

To ensure that the vector potential $\mathbf{A}$ corresponds to the specific classical wave-packet under the photon coherent state $| \phi \rangle_{\rm ph}$, the classical wavepacket condition must be imposed, requiring:
\begin{align}
\label{eq:A_cl}
    & \langle \phi |_{\rm ph} \mathbf{A} (\mathbf{r}, t) | \phi \rangle_{\rm ph}  =\mathbf{A}_{\rm cl} (\mathbf{r}, t)  \notag ,\\
    &  \mathbf{A}_{\rm cl} (\mathbf{r}, t) \equiv \sqrt{\frac{\hbar}{2\bm{\eps}_0 \omega_{{\rm in}} V}} 
    s_{\bm{\eps}}\left(t -  \frac{\hat{\mathbf{p}}_{\rm in} \cdot (\mathbf{r} - \mathbf{r}_0)}{c}\right)
    e^{-i \omega_{\rm in} \bigl( t -  \frac{\hat{\mathbf{p}}_{\rm in} \cdot (\mathbf{r} - \mathbf{r}_0)}{c} \bigr)} \bm{\eps} \nonumber\\ 
     & \quad \quad \quad  \quad + c.c. \,, 
\end{align}
where $\mathbf{A}_{\rm cl} (\mathbf{r}, t)$ is the vector potential for a classical wave packet, $\hat{\mathbf{p}}_{\rm in}=\mathbf{p}_{\rm in} / |\mathbf{p}_{\rm in}|$ is the unit vector along the incident photon propagation direction, $\omega_{\rm in}=|\mathbf{p}_{\rm in}|c$ is the incident photon's angular frequency. We set the location of the sample at the origin $\mathbf{r}_0 \equiv \mathbf{0}$ without loss of generality. 

To satisfy Eqs.\,\eqref{eq:Art}-\eqref{eq:A_cl} simultaneously, we require
\begin{align}\label{eq:stsq}
    &s_{\bm{\eps}}\left(t -  \frac{\hat{\mathbf{p}}_{\rm in} \cdot \mathbf{r}}{c}\right) 
    e^{-i \omega_{\rm in} \left( t -  \frac{\hat{\mathbf{p}}_{\rm in} \cdot \mathbf{r}}{c} \right)} 
    = \sum_{\mathbf{p}} \sqrt{\frac{\omega_{{\rm in}}}{\omega_{\mathbf{p}}}} e^{i (\mathbf{p} \cdot \mathbf{r} - \omega_{\mathbf{p}} t)} s_{\mathbf{p}\bm{\eps}} .
\end{align}
By following Eq. \eqref{eq:stsq} and converting the summation into an integral over photon momentum, we can establish the relationship between the real-time and momentum-space amplitudes as
\begin{align}\label{seps}
    s_{\bm{\eps}}(t) = V \int \frac{d^3 \mathbf{p}}{(2\pi)^3} \sqrt{\frac{\omega_{{\rm in}}}{\omega_{\mathbf{p}}}} e^{i (\mathbf{p} - \mathbf{p}_{\rm in}) \cdot \mathbf{r}  - i (\omega_{\mathbf{p}} - \omega_{\rm in}) t} s_{\mathbf{p}\bm{\eps}}.
\end{align}
Assume the X-ray beam has the cross-sectional area of $\sigma_{\rm beam}$ and that $s_{\bm{\eps}}(t)$ is slowly varying compared to the frequency of light, Eq.\,\eqref{seps} can be inverted as
\begin{align}
    {s}_{\mathbf{p}\bm{\eps}} &= \frac{\delta_{\mathbf{p}, (\mathbf{p}\cdot \hat{\mathbf{p}}_{\rm in})\hat{\mathbf{p}}_{\rm in}}}{V} \sigma_{\rm beam}\int_{-\infty}^{\infty} (c dt) \, s_{\bm{\eps}}(t) e^{i (\omega_{\mathbf{p}}  - \omega_{\rm in}) t},
\end{align}
where 
$\delta_{\mathbf{p}, (\mathbf{p}\cdot \hat{\mathbf{p}}_{\rm in})\hat{\mathbf{p}}_{\rm in}}$ ensures that $\mathbf{p}$ and $\hat{\mathbf{p}}_{\rm in}$ are parallel. Inserting this expression back into Eq.\,\eqref{eq:coherent}, we have constructed the generic coherent photon state with any possible real-time amplitude $s_{\bm{\eps}}(t)$. For synchrotron radiation in a continuous mode, the real-time amplitude for the photon beam with the polarization vector $\eps_{\rm in}$ can be expressed as  
\begin{align}
    s_{\bm{\eps}}^{\rm cont}(t) = s_0 \delta_{\bm{\eps},\bm{\eps}_{\rm in}}.
\end{align}
For a free-electron laser, the amplitude of the pulse train can be represented as a sum of multiple Gaussian pulses, each with a pulse width $\sigma_{\rm pr}$ and separated by a time interval $T_{\rm g}$ between consecutive pulses,
\begin{align}\label{eq:pulse-shape}
    s_{\bm{\eps}}^{\rm pulse}(t) &= \sum_{j = 0}^{n-1} \frac{s_0}{\sqrt{2 \pi \sigma_{\rm pr}^2}} e^{- \frac{(t - j T_{\rm g})^2}{2 \sigma_{\rm{ pr}}^2}} \delta_{\bm{\eps},\bm{\eps}_{\rm in}}.  
\end{align}
The $\sigma_{\rm{ pr}}$ and $T_{\rm g}$ are illustrated in Fig.\,\ref{fig:setupHPT}(a). The experiment commences at $t=0$ when the center of the first pulse hits the sample. The time interval between two consecutive pulses, $T_{\rm g}$, is usually much larger than the probe width $\sigma_{\rm pr}$, making the generated pulses well-separated. A realistic set of parameters can be referenced from the current EuXFEL setup \cite{kujala2020hard}, where $n = 2700$, $\sigma_{\rm pr} \sim 100$ fs, and $T_{\rm g} = 220$ ns, which satisfy the slow-varying $s(t)$ condition $\frac{1}{\omega_{\rm in}} \ll \sigma_{\rm pr}$ at X-ray frequencies.

\subsection{Single-Photon Scattering}
To analyze the detector responses of the incoming coherent X-ray pulses, we first study the single-photon scattering using the time-evolution operators under the electron-photon interaction. The single-photon scattering encompasses processes like diffraction, as well as non-resonant elastic and inelastic scattering. We assume a weak light-matter interaction, i.e., $\mathcal{H}_{\rm int} \ll \mathcal{H}_{\rm 0}$. As a result, one can expand the full time-evolution operator perturbatively as: 
\begin{align}
	&U(t, t_0)  = U_0(t, t_0)  -\frac{i}{\hbar}\int_{t_0}^{t} dt_1  U_0(t, t_1)\mathcal{H}_{\rm int}U_0(t_1, t_0) \nonumber\\ & \!\!\!\!\!\!+ \left(\frac{-i}{\hbar} \right)^2\int_{t_0}^{t} dt_1 \int_{t_0}^{t_1} dt_2 U_0(t, t_1) \mathcal{H}_{\rm int} U_0(t_1, t_2) \mathcal{H}_{\rm int}U_0(t_2, t_0)  \nonumber
 \\& \!\!\!\!\!\!+ \cdots , 
 \end{align}
where $U(t, t') = \exp\left(- \frac{i \mathcal{H} (t - t')}{\hbar} \right)$ is the time-evolution operator for the full Hilbert space, and $U_0(t, t') = \exp \left(- \frac{i \mathcal{H}_0 (t - t')}{\hbar} \right)$ is the time-evolution operator without electron-photon interaction. The order of perturbation can be interpreted as the number of times the photon scatters off the sample. 

The perturbatively-expanded time-evolution operator allows us to analyze the single-photon scattering under the electron-photon interaction. First, we define the scattered coherent photon intensity using photon number operator in the second-quantized form as 
\begin{align}
    \label{eq:intensity}
	I_{\mathbf{q}_1\bm{\eps}_1}(t) =  a^\dagger_{\mathbf{p}_1\bm{\eps}_1}(t) a_{\mathbf{p}_1\bm{\eps}_1}(t),
\end{align}
where $I_{\mathbf{q}_1\bm{\eps}_1}$ represents the intensity of the scattered light with momentum $\mathbf{p}_1$, polarization $\bm{\eps}_1$, and momentum transfers $\mathbf{q}_1 = \mathbf{p}_{\rm in} - \mathbf{p}_1$ from the incoming photon at momentum $\mathbf{p}_{\rm in}$ and polarization $\eps_{\rm in}$. The time-dependence of the operator in Eq. \eqref{eq:intensity} is defined under the full Heisenberg picture $\mathcal{O}(t) = e^{i \mathcal{H}t/\hbar}\mathcal{O} e^{-i \mathcal{H}t/\hbar}$. Consider the background scattering pattern of XPCS at time $t$, which is the quantum ensemble average of the scattered coherent photon intensity Eq.\,\eqref{eq:intensity}:
\begin{align}
	\braket{I_{\mathbf{q}_1\bm{\eps}_1}(t)} &=   \langle a^\dagger_{\mathbf{p}_1\bm{\eps}_1}(t) a_{\mathbf{p}_1\bm{\eps}_1}(t)\rangle,
\end{align}

Note that since we are computing the expectation value with respect to the electronic thermal states and photon coherent states, the first non-zero contribution to the intensity corresponds to the second-order response,
\begin{align}
\braket{I_{\mathbf{q}_1\bm{\eps}_1}(t
)} &=  
\langle \wick{\c1 U(t_0, t)\c1 a^\dagger_{\mathbf{p}_1\bm{\eps}_1} \c2 a_{\mathbf{p}_1\bm{\eps}_1}\c2 U(t, t_0)} \rangle,
\end{align} 
where the top lines denote the contractions between the creation and annihilation operators, specifically indicating that a creation operator in the time-evolution operator contracts with an annihilation operator in the intensity operator, and vice versa. The intensity can then be expressed in the form of the electron multipole-multipole correlation,
\begin{align}
    \braket{I_{\mathbf{q}_1\bm{\eps}_1}(t)}  &=  \frac{1}{\hbar^2} 
    \int_{t_0}^{t} \int_{t_0}^{t} 
    dt_1 dt_2   s^*_{\bm{\eps}_{\rm in}}(t_1) s_{\bm{\eps}_{\rm in}}(t_2)\langle \mathcal{D}_{\mathbf{q_1}}^\dagger(t_1) \mathcal{D}_{\mathbf{q_1}}(t_2)\rangle_{\rm e}
    \nonumber \\  & \quad \quad\quad\quad\quad  \times
    e^{-i (\omega_{\rm in} - \omega_{\mathbf{p}_1}) (t_2 - t_1)},\\
    \mathcal{D}_{\mathbf{q_1}}^{\bm{\eps}_{\rm in}\bm{\eps}_1}(t_{i}) &= \sum_{\mathbf{k}\alpha\beta} M_{\mathbf{k},\mathbf{p}_{\rm in},\mathbf{q}_1}^{\alpha\beta\bm{\eps}_{\rm in}\bm{\eps}_1} c^\dagger_{\mathbf{k}+\mathbf{q_1}, \alpha}(t_{i}) c_{\mathbf{k}, \beta}(t_{i}) \quad i = 1, 2 \, ,
\end{align}
where $\mathcal{D}$ is the electron multipole operator and $\langle...\rangle_{\rm e}$ represents the corresponding quantum state ensemble average but only for the electronic states. The detailed derivation can be found in Appendix\,\ref{Supp:single-photon}.

The electron multipole operator here is generic and could represent multiple different excitations, based on the various forms of the photon-electron scattering matrix element $M_{\mathbf{k},\mathbf{p}_{\rm in},\mathbf{q}_1}^{\alpha\beta\bm{\eps}_i\bm{\eps}_f}$. If we further assume the electron orbitals are plane waves in the
solid $\psi_{\kbf\alpha}(\rbf) = \frac{1}{\sqrt{V}}e^{i\kbf \cdot \rbf}$, then Eq. \eqref{eq:scatteringMatrixElement} is simplified as 
\begin{eqnarray}
M^{\alpha\beta\bm{\eps}_{i}\bm{\eps}_f}_{\kbf,\mathbf{p}_{\rm in},\mathbf{q}_1}
\!&=&\!\frac{e^2\hbar  \left(\bm{\eps}_f^* \cdot \bm{\eps}_{i}\right)\delta_{\alpha \beta}}{2m\bm{\eps}_0 V\sqrt{\omega_{\mathbf{p}_{i}}\omega_{\mathbf{p}_{i} - \mathbf{q}_1}}} \,,
\end{eqnarray}

As a result, the photon and electron momenta in the scattering matrix element are decoupled, and the multipole operator is simplified to be proportional to the electron number density operator:
\begin{equation}\label{sDapprox}
\begin{split}
    \mathcal{D}_{\mathbf{q_1}}^{\bm{\eps}_{\rm in}\bm{\eps}_1} &= \mathcal{M}_{\mathbf{p}_{\rm in},\mathbf{q}_1}^{\bm{\eps}_{\rm in}\bm{\eps}_1}\rho_{\mathbf{q}_1},\,\, \,\,\rho_{\mathbf{q}_1} =\sum_{\mathbf{k}} c^\dagger_{\mathbf{k}+\mathbf{q_1}}c_{\mathbf{k}},\\
    \mathcal{M}_{\mathbf{p}_{\rm in},\mathbf{q}_1}^{\bm{\eps}_{\rm in}\bm{\eps}_1} &=  \frac{e^2\hbar  \left(\bm{\eps}_1^* \cdot \bm{\eps}_{\rm in}\right)}{2m\bm{\eps}_0 V\sqrt{\omega_{\mathbf{p}_{\rm in}}\omega_{\mathbf{p}_{\rm in} - \mathbf{q}_1}}}, 
\end{split}
\end{equation}
where $\rho_{\mathbf{q}}$ represents the electron number density operator without band or spin indices. As a result, the single photon scattering intensity can be expressed as the electron density-density correlation function:
\begin{align} \label{single}
   &\braket{I_{\mathbf{q}_1 \bm{\eps}_1}(t)} = \frac{|\mathcal{M}_{\mathbf{p}_{\rm in},\mathbf{q_1}}^{\bm{\eps}_{\rm in} \bm{\eps}_1}|^2}{\hbar^2}\int_{t_0}^{t} \int_{t_0}^{t} 
    dt_1 dt_2 \langle \rho_{\mathbf{q_1}}^\dagger(t_1) \rho_{\mathbf{q_1}}(t_2)\rangle_{\rm e}
    \nonumber \\  & \quad \times s^*_{\bm{\eps}_{\rm in}}(t_1) s_{\bm{\eps}_{\rm in}}(t_2)
    e^{-i (\omega_{\rm in} - \omega_{\mathbf{p}_1}) (t_2 - t_1)}.
\end{align}
This exactly reproduces the results of time-resolved inelastic X-ray scattering, where similar calculations have been performed in various time-resolved spectroscopy works \cite{freericks2009theoretical,wang2018theory,freericks2018nonresonant,chen2019theory,wang2020time}.

\subsection{From Classical to Quantum XPCS: Four Configurations}

\begin{figure}
    \centering
\includegraphics[width = 1.00\linewidth]{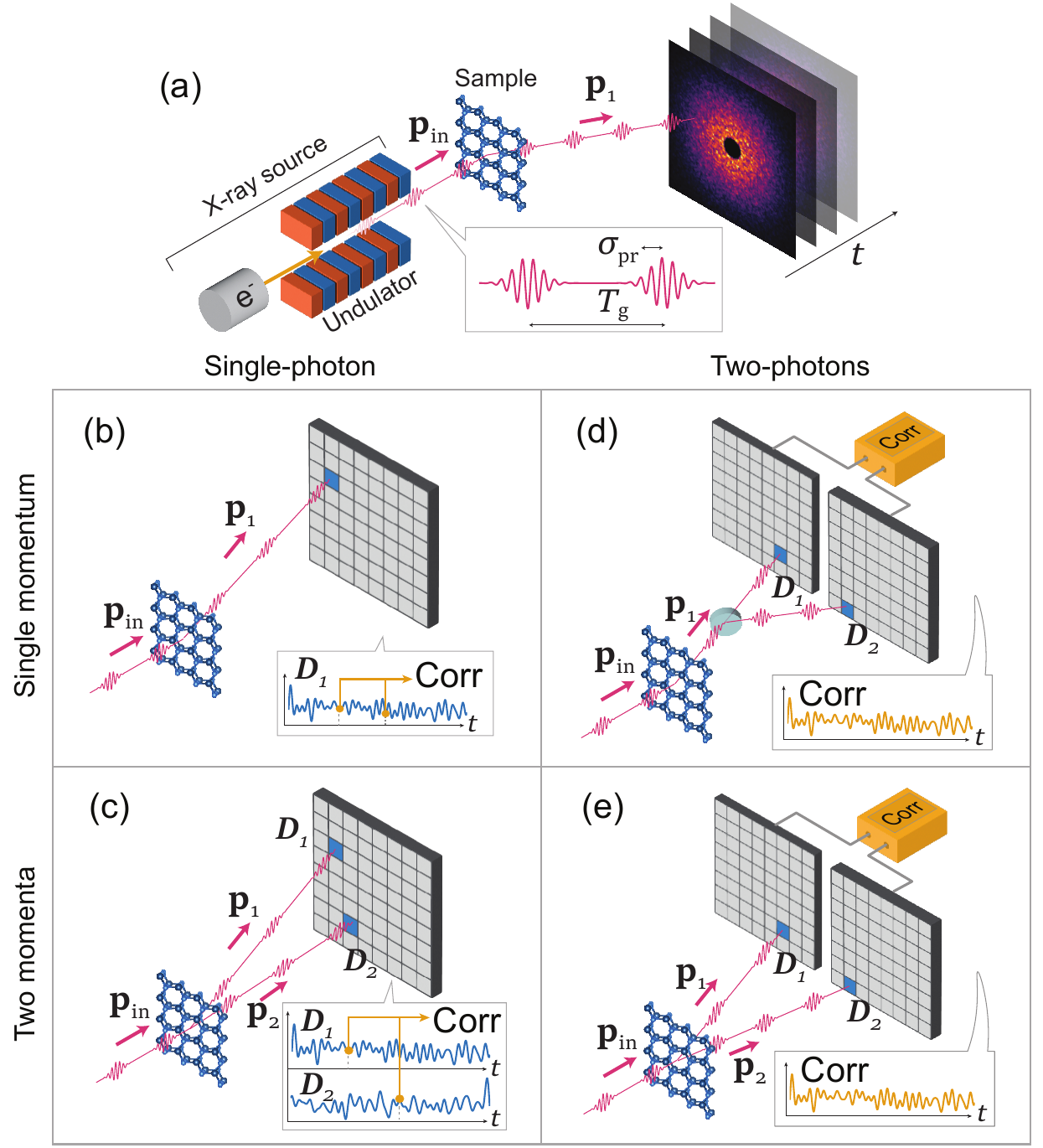}
    \caption{Proposed four XPCS configurations. Here $\mathbf{p}_{\rm{in}}$, $\mathbf{p}_1$, and $\mathbf{p}_2$ denote the incoming photon momenta, the scattered photon momentum received by detector $D_1$, and the scattered photon momentum received by detector $D_2$, respectively. $\sigma_{\rm pr}$ represents the X-ray pulse width in time domain, while $T_{\rm g}$ represents the time interval between two consecutive pulses. (a) The schematic of a general photon scattering process off a sample. (b-c) The conventional ``classical'' XPCS setup. The intensity of a scattered photon is measured as a function of time. Then, the digital correlation is performed for two scattering intensities sharing the same momentum (b) or having different scattering photon momenta (c). (d-e) The two HBT-inspired ``quantum'' XPCS. (d) The light travels in different optical paths before arriving at detectors $D_1$ and $D_2$. Then, the autocorrelator returns a signal only when the photon hits both detectors. (e) The two-photon, two-momenta quantum XPCS configuration. The configurations shown in figures (b) -- (e) correspond to the measurables in Eqs. \eqref{eq:twopoint1} -- \eqref{eq:twopoint4}, respectively.}
    \label{fig:setupHPT}
\end{figure}

Unlike single-photon scattering, the second-order photon correlation offers a unique opportunity to probe electronic dynamics in greater detail by granting access to higher-order electron correlations, particularly fourth-order electron correlations. In this section, we discuss the possibility of four different XPCS configurations (Figs.\,\ref{fig:setupHPT}(b)-\ref{fig:setupHPT}(e)) and their corresponding observables, which are the second-order photon correlation functions (Eqs.\,\eqref{eq:twopoint1}-\eqref{eq:twopoint4}) and further linked to fourth-order electron correlations. The first two setups (Figs.\,\ref{fig:setupHPT}(b) and \ref{fig:setupHPT}(c)) represent the conventional ``classical'' XPCS configurations, where scattered single-photon intensities are measured at different times. The observable, namely the second-order photon correlation function, is essentially the autocorrelation of the single-photon intensities, which is digitally obtained from single-photon intensity measurements. The key difference between Fig.\,\ref{fig:setupHPT}(b) and Fig.\,\ref{fig:setupHPT}(c) is whether the autocorrelation of photon intensity is performed at the same photon momentum transfer (as in Fig.\,\ref{fig:setupHPT}(b)) or at two different momentum transfers (as in Fig.\,\ref{fig:setupHPT}(c)). It is worth noting that, despite the technical feasibility of the configuration in Fig.\,\ref{fig:setupHPT}(c), current XPCS setups are still limited to the case in Fig.\,\ref{fig:setupHPT}(b), possibly due to limited motivation to explore the effects of two different momenta or limited data storage infrastructures.

In addition to the two ``classical'' XPCS configurations, here we propose two ``quantum'' XPCS configurations. These two configurations are inspired by the HBT experiments in quantum optics that measure the two-photon correlations but with finite photon momentum transfer, as shown in Fig.\,\ref{fig:setupHPT}(d) and \,\ref{fig:setupHPT}(e). Since two photons are observed, the observable of the second-order photon correlation function is no longer the single-photon intensity autocorrelation. Fig. \,\ref{fig:setupHPT}(d) corresponds to the case of photon splitting with a beam splitter, which allows one photon to be detected by one detector ($D_1$) while the other photon detected by the other detector ($D_2$). The two photons share the same scattering momentum $\mathbf{p}_1$. On the other hand, Fig. \,\ref{fig:setupHPT}(e) corresponds to the most-general case without a beam splitter, where the two photons with different momenta $\mathbf{p}_1$ and $\mathbf{p}_2$ are scattered into two detectors, respectively, and a correlation signal is generated when each of the two photons hits each of the two detectors simultaneously. The physical correlation between the detection events at the two detectors is measured directly. The finite area of the detectors with multiple pixels may appear to blur the distinction between the case in Fig.\,\ref{fig:setupHPT}(c) and Fig.\,\ref{fig:setupHPT}(e). However, they are fundamentally different: the ``classical'' two-momentum XPCS in Fig.\,\ref{fig:setupHPT}(c) registers the intensity directly and autocorrelation is computed after the measurements, while the quantum two-momenta XPCS Fig.\,\ref{fig:setupHPT}(e) directly captures the physical correlation between two scattered photons.

The four experimental configurations of XPCS in Fig. \,\ref{fig:setupHPT} provide an intuitive foundation to help construct the corresponding observable of second-order photon correlation functions. For the two classical XPCS configurations, the observable can be derived by following the standard procedure used to calculate the symmetric (real part) spectral density \cite{clerk2010introduction} of the single-photon intensity-intensity correlation, as summarized in Eqs.\,\eqref{eq:twopoint1} and \,\eqref{eq:twopoint2}. Notice that the ensemble average $\langle...\rangle$ is performed over the full Hamiltonian, including both photon and electron states. For completeness, the observables for the two quantum XPCS configurations are summarized upfront as in Eqs.\,\eqref{eq:twopoint3} and \,\eqref{eq:twopoint4}, with detailed derivation provided later on:

    \begin{align}
     &G^{(2)}_{\mathbf{q}_1\bm{\eps}_1}(t,t')_{\rm cl} 
     =  
       \mathrm{Re} \langle  I_{\mathbf{q}_1\bm{\eps}_1}(t) I_{\mathbf{q}_1\bm{\eps}_1}(t')\rangle =
       \nonumber \\ 
       & \mathrm{Re} \left\langle  a^\dagger_{\mathbf{p}_1\bm{\eps}_1}(t)a_{\mathbf{p}_1\bm{\eps}_1}(t)  
     a^\dagger_{\mathbf{p}_1\bm{\eps}_1}(t') a_{\mathbf{p}_1\bm{\eps}_1}(t')\right\rangle  \label{eq:twopoint1},\\ \nonumber \\
    &G^{(2)}_{\mathbf{q}_1 \bm{\eps}_1 \mathbf{q}_2 \bm{\eps}_2}(t,t')_{\rm cl} = \mathrm{Re} \langle I_{\mathbf{q}_1}(t) I_{\mathbf{q}_2}(t')\rangle  =  
    \nonumber \\ &
      \mathrm{Re} \left\langle  a^\dagger_{\mathbf{p}_1\bm{\eps}_1}(t)a_{\mathbf{p}_1\bm{\eps}_1}(t)  
    a^\dagger_{\mathbf{p}_2\bm{\eps}_2}(t') a_{\mathbf{p}_2\bm{\eps}_2}(t')\right\rangle  \label{eq:twopoint2}, 
    \\ \nonumber \\
&G^{(2)}_{\mathbf{q}_1 \bm{\eps}_1}(t,t') = \nonumber \\
    &\left
    \langle \tilde{T}\{a^\dagger_{\mathbf{p}_1\bm{\eps}_1}(t) a^\dagger_{\mathbf{p}_1\bm{\eps}_1}
    (t')
    \}T\{a_{\mathbf{p}_1\bm{\eps}_1}(t')  a_{\mathbf{p}_1\bm{\eps}_1} (t)\}\right\rangle\label{eq:twopoint3},
    \\ \nonumber \\
   &G^{(2)}_{\mathbf{q}_1 \bm{\eps}_1 \mathbf{q}_2 \bm{\eps}_2}(t,t')  = 
   \nonumber\\&
   \left\langle \tilde{T}\{a^\dagger_{\mathbf{p}_1\bm{\eps}_1}(t) a^\dagger_{\mathbf{p}_2\bm{\eps}_2}
    (t') 
    \}T\{ a_{\mathbf{p}_2\bm{\eps}_2}(t')  a_{\mathbf{p}_1\bm{\eps}_1} (t)\}\right\rangle\label{eq:twopoint4}, 
\end{align}

\noindent where $\mathbf{p}_{1}$ ($\mathbf{p}_{2}$) is the momentum of the first (second) scattered photon with polarization $\bm{\eps}_1$($\bm{\eps}_2$), and $\mathbf{q}_{1}=\mathbf{p}_{\rm{in}}-\mathbf{p}_{1}$ ($\mathbf{q}_{2}=\mathbf{p}_{\rm{in}}-\mathbf{p}_{2}$) is the corresponding momentum transfer. In Eqs.\,\eqref{eq:twopoint3} and \,\eqref{eq:twopoint4}, $T$ ($\Tilde{T}$) is the time (anti-time) ordering operator, which guarantees the acted operators are in time (anti-time) order.

Before proceeding with the derivation of the main results of Eqs.\,\eqref{eq:twopoint3} and \,\eqref{eq:twopoint4}, it is important to briefly digress on a key point that requires attention: the XPCS measurable also depends on the timescale of the dynamics. For ultrafast dynamics where the characteristic time scale is much shorter than the detection time scale, the electron configurations in phase space are fully spanned during the measurement period, and when using time-average to approximate ensemble average, we actually have  $\langle I_{\mathbf{q}_1}(t) I_{\mathbf{q}_1}(t')\rangle \approx \langle I_{\mathbf{q}_1}(t) \rangle \langle I_{\mathbf{q}_1}(t')\rangle$, and no photon correlation effect is measured. Only for slow dynamics can we sample one electron configuration in the phase space and measure the sampled eigenvalue of the operator $I_{\mathbf{q}_1}(t)$. This is the reason that conventional XPCS is used to study slow dynamics, and the time average is used as a substitute for the ensemble average to measure $\langle I_{\mathbf{q}_1}(t) I_{\mathbf{q}_1}(t') \rangle$, assuming that the system is thermalized. Therefore, special attention shall be paid when using XPCS to study ultrafast dynamics or to investigate exotic quantum systems that break ergodicity.

Now we derive the observable of second-order photon correlation function for quantum XPCS setups described in Figs.\,\ref{fig:setupHPT}(d) and \ref{fig:setupHPT}(e), which enables us to measure the two-photon correlation without the single-photon double-scattering events. 

Intuitively, these observables are directly related to the probability that the photon states evolve over time under the full Hamiltonian and are eventually detected by the two detectors $D_1$ and $D_2$, i.e.,  
\begin{align}\label{eq:G2_intuitive}
    G^{(2)} (t, t') \sim |\braket{D_1(t)D_2(t')|\Psi_0}|^2,
\end{align}
where $\ket{\Psi_0}$ is the initial state of the full system defined from Sec.\,\ref{sec:photon}. It is worth mentioning that the expression in Eq.\,\eqref{eq:G2_intuitive} in the ``quantum'' XPCS setup is no longer the intensity auto-correlation function. Even so, for simplicity and without causing confusion, the observable is still referred to as $G^{(2)}$.
In addition, $t$ and $t'$ are defined as the time when each of the two photons hit the sample. At time $\tilde{t}$, the correlator (shown in Fig. \ref{fig:setupHPT}(e)) would measure the correlation between the photons which interacted with the sample at two earlier times $t = \tilde{t} - L_1/c$ and $t' = \tilde{t} - L_2/c$, respectively, where $L_1$ and $L_2$ are the photon path distance from the sample to the detector. Notice that if we assume $L_1 > L_2$, it induces the time lag $\frac{L_1 - L_2}{c}$ such that the photon arrives at $D_2$ earlier than $D_1$, or, say, $t$ is earlier than $t'$. 

To rigorously obtain an expression of the quantum XPCS observables, we need to construct the observed photon state by the two detectors $\ket{D_1(t)D_2(t')}$, bearing in mind that the times $t$ and $t'$ are defined as photon arrival times at the sample---not the detector.

Note that measuring these two photons does not uniquely specify the final state; all we can say is the final state consists of at least two photons. The most general such state can be constructed by acting with two-photon creation-operators on some arbitrary quantum state $\ket{\Phi}$. This state $\ket{\Phi}$ accounts for all of the other scattered photons that are not relevant to our measurement, and the state also satisfies the completeness relation $\sum_{\Phi} \ket{\Phi} \bra{\Phi} = 1$. A natural choice for the final state is therefore $a^\dagger_{\mathbf{p}_1\bm{\eps}_1}(t)a^\dagger_{\mathbf{p}_2\bm{\eps}_2}(t')\ket{\Phi}$.  Unfortunately, there is an ambiguity in defining the final state in this way as creation operators at different times need not commute, so we could equally well take $a^\dagger_{\mathbf{p}_2\bm{\eps}_2}(t')a^\dagger_{\mathbf{p}_1\bm{\eps}_1}(t)\ket{\Phi}$ as our final state. How can we determine the correct operator ordering? While the detectors technically annihilate the two photons simultaneously, the differing path lengths mean that each photon leaves the sample at a different time, so the detectors are \textit{in effect} measuring them sequentially in the time order. Therefore, the composite two-photon annihilation operator is time-ordered, $T\{a_{\mathbf{p}_1\bm{\eps}_1}(t)a_{\mathbf{p}_2\bm{\eps}_2}(t')\}$, so the final state is given by

\begin{equation}
\begin{split}
    \bra{D_1(t)D_2(t')} = \bra{\Phi} T\{a_{\mathbf{p}_1\bm{\eps}_1}(t)a_{\mathbf{p}_2\bm{\eps}_2}(t')\},\\\ket{D_1(t)D_2(t')} =\left(T\{a_{\mathbf{p}_1\bm{\eps}_1}(t)a_{\mathbf{p}_2\bm{\eps}_2}(t')\}\right)^{\dagger} \ket{\Phi}.
\end{split}
\end{equation}
The state $\ket{D_1(t)D_2(t')}$ represents the most general two-photon quantum state consisting of one photon with momentum $\mathbf{p}_1$ and polarization $\bm{\eps}_1$ at time $t$ and the other photon of momentum $ \mathbf{p}_2$ and polarization $\bm{\eps}_2$ at time $t'$. Note that this state $\ket{D_1(t)D_2(t')}$ is defined in the Heisenberg picture and thus does not evolve with time. The labels $t,t'$ indicate the times at which the photons scatter off the sample.

Putting everything together for an initial state $\ket{\Psi_0}$ and final two-photon state $\ket{D_1(t)D_2(t')}$, the observable---namely the two-photon second-order correlation function---can be written by summing over all the unmeasured photon states $\Phi$:  

\begin{align}
    &G^{(2)}_{\mathbf{q}_1\bm{\eps}_1\mathbf{q}_2\bm{\eps}_2} (t, t')
     = \sum_\Phi \langle \Psi_0|\left(T\{a_{\mathbf{p}_1\bm{\eps}_1}(t)a_{\mathbf{p}_2\bm{\eps}_2}(t')\}\right)^\dagger|\Phi\rangle \nonumber \\ &\quad\quad\quad\quad\quad\quad\quad\times \langle \Phi | T\{a_{\mathbf{p}_1\bm{\eps}_1}(t)a_{\mathbf{p}_2\bm{\eps}_2}(t') \}|\Psi_0\rangle 
     \nonumber  \\&=  \langle \Psi_0|
    \tilde{T} \{a^{\dagger}_{\mathbf{p}_1\bm{\eps}_1}(t) a^{\dagger}_{\mathbf{p}_2\bm{\eps}_2}(t') \}T\{a_{\mathbf{p}_2\bm{\eps}_2}(t') a_{\mathbf{p}_1\bm{\eps}_1}(t)\} |\Psi_0\rangle.
\end{align}
The polarization states of the two photons can be summed over if the detectors are not sensitive to polarization. 

By generalizing the initial pure state $\ket{\Psi_0}$ to a mixed state $\rho = \frac{e^{- \beta \mathcal{H}_{\rm e}}}{Z} \otimes |{\phi}\rangle_{\rm ph}\langle{\phi}|_{\rm ph}$, where $Z$ is the corresponding partition function, Eq.\,\eqref{eq:twopoint4} is obtained as the measurable of quantum XPCS.

From a theoretical perspective, despite measuring the second-order photon correlation function, the key difference between the classical XPCS and quantum XPCS lies in the ordering of the photon creation and annihilation operators. In the classical limit where the intensity operator commutes, the classical XPCS and quantum XPCS give the same expression. In classical XPCS, the second-order correlation comes from the second-order autocorrelation of the single-photon intensity operator. On the contrary, the observables in the quantum XPCS setups, as shown in  Eqs.\,\eqref{eq:twopoint3} and \eqref{eq:twopoint4}, only contain the two-photon correlation with the second-order single-photon contribution from Eqs.\,\eqref{eq:twopoint1} and \eqref{eq:twopoint2} excluded. This gives the quantum XPCS a huge advantage in studying higher-order correlation functions. Therefore, we focus primarily on the quantum XPCS setups in the following.

\subsection{From Two-Photon Correlation to Fourth-Order Electron Density Correlations}

In this section, we establish the connection between the measured two-photon correlation function in Eq.\,\eqref{eq:twopoint4} and the intrinsic properties of electron correlation function using perturbation theory. The first non-zero term of the perturbation is the fourth-order electron correlation, which can be represented by Wick's contraction as 
\begin{align}\label{eq:wick4}
    & G^{(2)}_{\mathbf{q}_1\bm{\eps}_1\mathbf{q}_2\bm{\eps}_2} (t, t') = \nonumber \\
    & \sum_{\bm{\eps}_1\bm{\eps}_2} \wick{\langle \c1 U(t_0, t) \c1 a^{\dagger}_{\mathbf{p}_1\bm{\eps}_1} \c2 U(t, t')
    \c2 a^{\dagger}_{\mathbf{p}_2\bm{\eps}_2} \c3 a_{\mathbf{p}_2\bm{\eps}_2} \c3 U(t', t) \c4 a_{\mathbf{p}_1\bm{\eps}_1}\c4 U(t, t_0) \rangle}\nonumber \\
     &+ \wick{\langle  \c2 U \c1 (t_0, t)  \c1 a^{\dagger}_{\mathbf{p}_1\bm{\eps}_1}  U(t, t')
    \c2 a^{\dagger}_{\mathbf{p}_2\bm{\eps}_2} \c3 a_{\mathbf{p}_2\bm{\eps}_2} \c3 U(t', t) \c4 a_{\mathbf{p}_1\bm{\eps}_1}\c4 U(t, t_0) \rangle} \nonumber \\
     &+ \wick{\langle \c1 U(t_0, t) \c1 a^{\dagger}_{\mathbf{p}_1\bm{\eps}_1} \c2 U(t, t') 
    \c2 a^{\dagger}_{\mathbf{p}_2\bm{\eps}_2} \c3 a_{\mathbf{p}_2\bm{\eps}_2}  U(t', t) \c4 a_{\mathbf{p}_1\bm{\eps}_1} \c4 U \c3(t, t_0) \rangle} 
     \nonumber \\ 
    &+ \wick{\langle \c2 U \c1 (t_0, t)  \c1 a^{\dagger}_{\mathbf{p}_1\bm{\eps}_1}  U(t, t')
    \c2 a^{\dagger}_{\mathbf{p}_2\bm{\eps}_2} \c3 a_{\mathbf{p}_2\bm{\eps}_2}  U(t', t) \c4 a_{\mathbf{p}_1\bm{\eps}_1} \c4 U \c3(t, t_0) \rangle} .
\end{align}
This perturbative expansion corresponds to the two-photon scattering events as the contraction with $U(t_a,t_b)$ representing the scattering in the interval $[t_a, t_b]$. Through a similar calculation, which has been performed in the single-photon scattering, it can be shown that the $ G^{(2)}_{\mathbf{q}_1 \mathbf{q}_2} (t, t')$ measurement corresponds to the four-point electron density correlation function:

\begin{align}
\label{eq:G2_full}
    G^{(2)}_{\mathbf{q}_1\bm{\eps}_1\mathbf{q}_2\bm{\eps}_2} (t, t')
    &= \frac{1}{\hbar^4} \nonumber |\mathcal{M}_{\mathbf{p}_{\rm in},\mathbf{q}_1}^{\bm{\eps}_{\rm in}\bm{\eps}_{1}}|^2|\mathcal{M}_{\mathbf{p}_{\rm in},\mathbf{q}_2}^{\bm{\eps}_{\rm in}\bm{\eps}_{2}}|^2 \\ &\times 
    \int_{(t_1, t_2, t_3, t_4) \in \Omega \times \Omega} dt_1 dt_2 dt_3 dt_4 \nonumber\\
     & \times s^*_{\bm{\eps}_{\rm in}}(t_1) s^*_{\bm{\eps}_{\rm in}}(t_2) s_{\bm{\eps}_{\rm in}}(t_3) s_{\bm{\eps}_{\rm in}}(t_4)
      \nonumber \\& \times e^{- i (\omega_{\rm in} - \omega_{\mathbf{p}_1}) (t_4 -t_1)} e^{- i (\omega_{\rm in} - \omega_{\mathbf{p}_2}) (t_3 -t_2)}  
    \nonumber \\ & \times \langle \tilde{T} \{
    \rho_{\vb{q}_1}^\dagger (t_1) 
    \rho_{\vb{q}_2}^\dagger (t_2) \}T\{
    \rho_{\vb{q}_2}(t_3) 
    \rho_{\vb{q}_1}(t_4)\}\rangle_{\rm e},
\end{align} 
where $\Omega= \{(t_x, t_y):  \min(t_x, t_y) < t  ,  \max(t_x, t_y) < t' \}.$ The electron contribution is $C^{\rho\rho\rho\rho}_{\mathbf{q}_1\mathbf{q}_2}( t_1, t_2, t_3, t_4) = \langle \tilde{T} \{\rho_{\vb{q}_1}^\dagger (t_1) \rho_{\vb{q}_2}^\dagger (t_2) \}T\{\rho_{\vb{q}_2}(t_3) \rho_{\vb{q}_1}(t_4)\}\rangle_{\rm e}$, where $t_1$, $t_4$ are the time when the electron-photon interaction creates the electron-hole pair and is detected by detector $D_1$ after $t$. Similarly, $t_2$ and $t_3$ are the time to create the electron-hole pair which is detected by detector $D_2$ after $t'$.

\section{Breakdown and generalization of the Siegert relation}\label{sec:siegert}

\begin{figure}
     \centering
     \includegraphics[width = 1\linewidth]{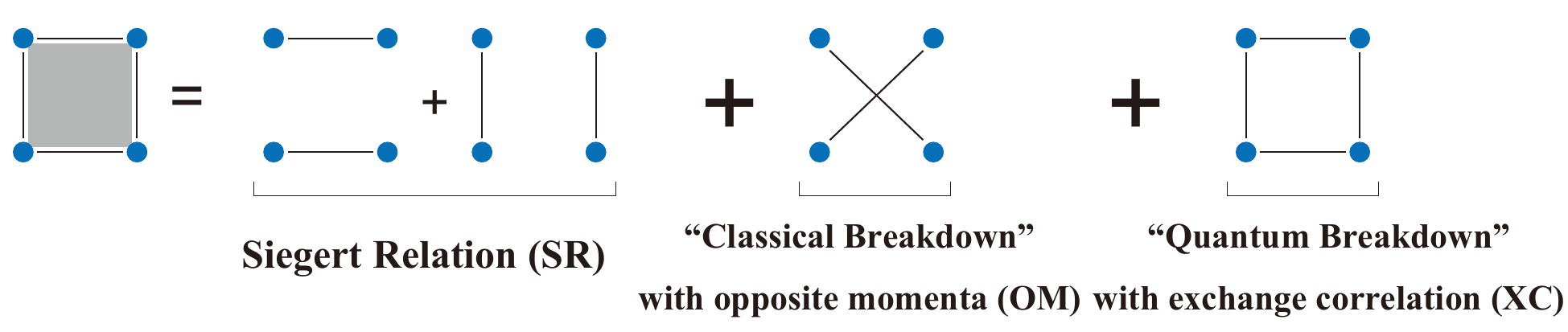}
     \caption{Schematic of the breakdown of the Siegert relation. The opposite momenta (OM) contribution in a system without spatial translational symmetry will cause the ``classical breakdown'' of the original Siegert relation. The exchange correlation (XC) contribution will cause the ``quantum breakdown'' of the Siegert relation in a quantum electronic system with exchange-correlation effects.}
     \label{fig:breakdown}
 \end{figure}

To link the four-point electron density correlation to the Siegert relation, we decompose the correlation function $C^{\rho\rho\rho\rho}_{\mathbf{q}_1\mathbf{q}_2}( t_1, t_2, t_3, t_4)$ using cumulant expansion as follows (Fig. \ref{fig:breakdown}),
\begin{widetext}
\begin{align}C^{\rho\rho\rho\rho}_{\mathbf{q}_1\mathbf{q}_2}( t_1, t_2, t_3, t_4)&= C^{\rho\rho\rho\rho}_{\mathbf{q}_1\mathbf{q}_2,\rm SR}( t_1, t_2, t_3, t_4) + C^{\rho\rho\rho\rho}_{\mathbf{q}_1\mathbf{q}_2,\rm OM}( t_1, t_2, t_3, t_4) +  C^{\rho\rho\rho\rho}_{\mathbf{q}_1\mathbf{q}_2,\rm XC}( t_1, t_2, t_3, t_4) 
,\\
C^{\rho\rho\rho\rho}_{\mathbf{q}_1\mathbf{q}_2,\rm SR}( t_1, t_2, t_3, t_4)&= \braket{\rho_{\vb{q}_1}^\dagger (t_1)  \rho_{\vb{q}_2}(t_3) }_{\rm e}^{\rm c} \braket{ \rho_{\vb{q}_2}^\dagger (t_2)  \rho_{\vb{q}_1}(t_4)}_{\rm e}^{\rm c}+\braket{ \rho_{\vb{q}_1}^\dagger (t_1) \rho_{\vb{q}_1}(t_4)}_{\rm e}^{\rm c} \braket{ \rho_{\vb{q}_2}^\dagger (t_2)  \rho_{\vb{q}_2}(t_3) }_{\rm e}^{\rm c}
    ,\\
C^{\rho\rho\rho\rho}_{\mathbf{q}_1\mathbf{q}_2,\rm OM}( t_1, t_2, t_3, t_4)&=\braket{\tilde{T} \{ \rho_{\vb{q}_1}^\dagger (t_1) 
    \rho_{\vb{q}_2}^\dagger (t_2)\}}_{\rm e}^{\rm c}\braket{T\{\rho_{\vb{q}_2}(t_3) 
    \rho_{\vb{q}_1}(t_4)\}}_{\rm e}^{\rm c} 
    ,\\
    C^{\rho\rho\rho\rho}_{\mathbf{q}_1\mathbf{q}_2,\rm  XC}( t_1, t_2, t_3, t_4)&= \langle \tilde{T} \{
    \rho_{\vb{q}_1}^\dagger (t_1) 
    \rho_{\vb{q}_2}^\dagger (t_2) \}T\{
    \rho_{\vb{q}_2}(t_3) 
    \rho_{\vb{q}_1}(t_4) \}\rangle _{\rm e}^{\rm c},
\end{align}
\end{widetext}
where $\langle...\rangle^{\rm c}_{\rm e}$ represents the electronic state's quantum state ensemble average within the corresponding cumulant expanded component represented as ``$\rm{c}$''. $C^{\rho\rho\rho\rho}_{\mathbf{q}_1\mathbf{q}_2,\rm SR}$ is the contribution quantified by the Siegert relation (SR). $C^{\rho\rho\rho\rho}_{\mathbf{q}_1\mathbf{q}_2,\rm OM}$ represents the two-momentum contribution which requires opposite momenta (OM), $\mathbf{q}_1 = -\mathbf{q}_2$, to be non-zero in a system with spatial translational invariance. $C^{\rho\rho\rho\rho}_{\mathbf{q}_1\mathbf{q}_2,\rm XC}$ represents the exchange-correlation (XC) contributions from exchanging the operators. The first two terms (SR and OM) are generated solely by Wick's theorem, while the last term (XC) arises exclusively from the four-electron correlation.

To compare the time-domain correlation function with the Siegert relation, we can consider the simple two-pulse structure in Eq.\,\eqref{eq:pulse-shape} where $n = 2$, $T_{\rm g} = t' - t$, and $\sigma_{ \rm{pr}} \ll T_{\rm g}$, which gives the form of $ G^{(2)}_{\mathbf{q}_1\mathbf{q}_2}$ as 

\begin{align}
    &G^{(2)}_{\mathbf{q}_1\bm{\eps}_1\mathbf{q}_2\bm{\eps}_2} (t, t')\nonumber \\
    &= |\mathcal{M}_{\mathbf{p}_{\rm in},\mathbf{q}_1}^{\bm{\eps}_{\rm in}\bm{\eps}_{1}}|^2|\mathcal{M}_{\mathbf{p}_{\rm in},\mathbf{q}_2}^{\bm{\eps}_{\rm in}\bm{\eps}_{2}}|^2
    \sum_{t_i \in \{t, t'\}} \alpha^*(t_2, t_1) \alpha (t_3, t_4) 
    \nonumber \\ &\times 
    C^{\rho\rho\rho\rho}_{\mathbf{q}_1\mathbf{q}_2}(t_1, t_2, t_3, t_4), \\
    &\alpha(t_i, t_j)  = e^{-i (\omega_{\rm in} - \omega_{\vb{p}_1}) t_i} e^{-i (\omega_{\rm in} - \omega_{\vb{p}_2}) t_j} 
    \nonumber \\ & \times
    \begin{cases}
        \frac{1}{4} \quad t_i = t \quad t_j = t' \\
        \frac{1}{4} \quad t_i = t' \quad t_j = t \\
        \frac{3}{4} \quad t_i = t \quad t_j = t \\
        0 \quad t_i = t' \quad t_j = t'
    \end{cases}.
\end{align}
Here function $\alpha(t_i,t_j)$ is calculated from the integral in Eq. \eqref{eq:G2_full} with the contribution from the pulse shape function and the phase factor. From this section forward, we assume $\hbar = 1$ to simplify the mathematical expressions. Although this autocorrelation contains multiple terms non-existent in classical XPCS, we can focus solely on the contribution from the ``time-delay'' contribution  $C^{\rho\rho\rho\rho}_{\mathbf{q}_1\mathbf{q}_2}( t_1, t_2, t_2, t_1)$ with $t_2 > t_1$. This time-delay contribution of quantum XPCS is in close analogy to the classical XPCS case.

Before demonstrating how exchange-correlation leads to a breakdown of the Siegert relation, we first reproduce the Siegert relation using our current formalism and derive a generalized Siegert relation for two momenta. Under the classical limit where the electron density operator is not constrained by commutation relations and can be interchanged freely, the two-photon correlation can be written as 
\begin{align}
\label{eq:G2}
    G^{(2)}_{\mathbf{q}_1\bm{\eps}_1\mathbf{q}_2\bm{\eps}_2} (t, t') =&|\mathcal{M}_{\mathbf{p}_{\rm in},\mathbf{q}_1}^{\bm{\eps}_{\rm in} \bm{\eps}_{1}}|^2|\mathcal{M}_{\mathbf{p}_{\rm in},\mathbf{q}_2}^{\bm{\eps}_{\rm in} \bm{\eps}_{2}}|^2 C^{\rho\rho\rho\rho}_{\mathbf{q}_1\mathbf{q}_2}( t_1, t_2, t_2, t_1).
\end{align}
As a result, this classical result matches the time-delay component of the quantum correlation function.

Assume the system has the spatial translational symmetry, i.e. momentum conservation, we have $C^{\rho\rho\rho\rho}_{\mathbf{q}_1\mathbf{q}_2,\rm OM} = 0$ when $\mathbf{q}_1 = \mathbf{q}_2$. For classical charge systems without electron exchange-correlation effects, we also have $C^{\rho\rho\rho\rho}_{\mathbf{q}_1\mathbf{q}_2,\rm XC} = 0$. The electron density correlation then is expressed as:
\begin{equation}
\begin{split}
   &C^{\rho\rho\rho\rho}_{\mathbf{q}_1\mathbf{q}_1}( t_1, t_2, t_2, t_1) = C^{\rho\rho\rho\rho}_{\mathbf{q}_1\mathbf{q}_1,\rm SR} ( t_1, t_2, t_2, t_1)
   \\&=\langle \rho_{\mathbf{q}_1}^{\dagger}(t_2)\rho_{\mathbf{q}_1}(t_2) 
\rangle_{\rm e}^{\rm c}\langle \rho_{\mathbf{q}_1}^{\dagger}(t_1)\rho_{\mathbf{q}_1}(t_1)  
\rangle_{\rm e}^{\rm c}
   \\&+\langle \rho_{\mathbf{q}_1}^{\dagger}(t_1)\rho_{\mathbf{q}_1}(t_2) 
\rangle_{\rm e}^{\rm c}\langle \rho_{\mathbf{q}_1 }^{\dagger}(t_2)\rho_{\mathbf{q}_1}(t_1)  
\rangle_{\rm e}^{\rm c}.
\end{split}
\end{equation}
This expression can be derived from the Eq.\,\eqref{eq:wick4} based on Wick's theorem or equivalently derived from Isserlis' theorem \cite{isserlis1918formula} by assuming classical Gaussian distribution under certain fluctuations of the scattering electron density.

For the system with time-translational symmetry, we have the following equal-time correlations $ \langle \rho_{\mathbf{q}_1}^{\dagger}(t_1)\rho_{\mathbf{q}_1}(t_1)  \rangle_{\rm e}^{\rm c} = \langle \rho_{\mathbf{q}_1}^{\dagger}(t_2)\rho_{\mathbf{q}_1}(t_2)\rangle_{\rm e}^{\rm c}=\langle\rho_{\mathbf{q}_1}^{\dagger}(t)\rho_{\mathbf{q}_1}(t)\rangle_{\rm e}^{\rm c}$ for any time $t$. If we label the time difference as $t_2-t_1 = t$ and set $t_1 = 0$, we can define the normalized electron density correlation functions as 
\begin{equation}
\begin{split}
   g^{(1)}_{\mathbf{q}_1}(t) &= \frac{\langle\rho_{\mathbf{q}_1}^{\dagger}(0)\rho_{\mathbf{q}_1}(t)\rangle_{\rm e}^{\rm c}}{\langle \rho_{\mathbf{q}_1}^{\dagger}(t)\rho_{\mathbf{q}_1}(t) \rangle_{\rm e}^{\rm c}}, \\   g^{(2)}_{\mathbf{q}_1}(t) &=\frac{ C^{\rho\rho\rho\rho}_{\mathbf{q}_1\mathbf{q}_1}( 0, t, t, 0)}{ (\langle\rho_{\mathbf{q}_1}^{\dagger}(t)\rho_{\mathbf{q}_1}(t) \rangle_{\rm e}^{\rm c} )^2}.
\end{split}
\end{equation}
Then we finally reach the Siegert relation \cite{siegert1943fluctuations}, which is expressed as $C^{\rho\rho\rho\rho}_{\mathbf{q}_1\mathbf{q}_1} = C^{\rho\rho\rho\rho}_{\mathbf{q}_1\mathbf{q}_1,\rm SR}$, or in a more common form:
\begin{equation}\label{eq:SRorigin}
\begin{split}
   g^{(2)}_{\mathbf{q}_1}(t) &= 1 + \left|g^{(1)}_{\mathbf{q}_1}(t)\right|^2.
\end{split}
\end{equation}
Notice that this relation is valid only when the system is classical and conserves momentum with negligible quantum exchange-correlation contribution.

If we allow $\mathbf{q}_1  \neq \mathbf{q}_2$ based on our generalized XPCS setup, particularly the setup in Fig.\,\ref{fig:setupHPT}(e), the Siegert relation can actually be generalized into the two-momentum case. For the system with spatial translational symmetry, $\braket{ \rho_{\vb{q}_1}^\dagger (t_1) 
    \rho_{\vb{q}_2} (t_2)}_{\rm e}^{\rm c} = 0$ when $\vb{q}_1 \neq \vb{q}_2$, however, the only non-trivial case is when $\mathbf{q}_1  = \mathbf{q}_2$ and $\mathbf{q}_1  = -\mathbf{q}_2$. In the first case, the only non-zero expression $C^{\rho\rho\rho\rho}_{\mathbf{q}_1\mathbf{q}_2, \rm SR}$ represents the single momentum scattering. In the second case, the non-zero contribution from $C^{\rho\rho\rho\rho}_{\mathbf{q}_1\mathbf{q}_2,\rm OM}$ can be attributed to the excitation from the first photon scattering and the de-excitation from the second photon scattering, which also results in a single momentum scattering process.

If we neglect the exchange-correlation effects but keep the two momenta $\mathbf{q}_1$ and $\mathbf{q}_2$ general, we have the generalized form of the Siegert relation for the two-momentum case as $C^{\rho\rho\rho\rho}_{\mathbf{q}_1\mathbf{q}_2} = C^{\rho\rho\rho\rho}_{\mathbf{q}_1\mathbf{q}_2,\rm SR} + C^{\rho\rho\rho\rho}_{\mathbf{q}_1\mathbf{q}_2,\rm OM}$, or equivalently: 
\begin{align}
   g^{(2)}_{\mathbf{q}_1\mathbf{q}_2}(t) &= 1 + \left|g^{(1)}_{\mathbf{q}_1\mathbf{q}_2}(t)\right|^2+\left|\tilde{g}^{(1)}_{\mathbf{q}_1\mathbf{q}_2}(t)\right|^2 \label{eq:SRtwoMomenta},
\end{align}
where 
\begin{align}
    g^{(2)}_{\mathbf{q}_1\mathbf{q}_2}(t) &=\frac{ C^{\rho\rho\rho\rho}_{\mathbf{q}_1\mathbf{q}_2}( 0, t, t, 0)}{ \langle \rho_{\mathbf{q}_1}^{\dagger}(0)\rho_{\mathbf{q}_1}(0) \rangle^{\rm c}_{\rm e}\langle \rho_{\mathbf{q}_2}^{\dagger}(t)\rho_{\mathbf{q}_2}(t) \rangle^{\rm c}_{\rm e}},
\end{align}
and $g^{(1)}_{\mathbf{q}_1\mathbf{q}_2}(t) = \sqrt{\frac{\langle\rho_{\mathbf{q}_1}^{\dagger}(0)\rho_{\mathbf{q}_2}(t)\rangle^{\rm c}_{\rm e}\langle\rho_{\mathbf{q}_2}^{\dagger}(t)\rho_{\mathbf{q}_1}(0)\rangle^{\rm c}_{\rm e}}{\langle \rho_{\mathbf{q}_1}^{\dagger}(0)\rho_{\mathbf{q}_1}(0) \rangle^{\rm c}_{\rm e}\langle \rho_{\mathbf{q}_2}^{\dagger}(t)\rho_{\mathbf{q}_2}(t) \rangle^{\rm c}_{\rm e}}}$ represents the contribution from the Siegert relation term, and $\tilde{g}^{(1)}_{\mathbf{q}_1\mathbf{q}_2}(t)= \sqrt{\frac{\langle\tilde{T} \{ \rho_{\vb{q}_1}^\dagger (0) 
    \rho_{\vb{q}_2}^\dagger (t)\}\rangle_{\rm e}^{\rm c}\langle T\{\rho_{\vb{q}_2}(t) 
        \rho_{\vb{q}_1}(0)\}\rangle_{\rm e}^{\rm c}}{\langle \rho_{\mathbf{q}_1}^{\dagger}(0)\rho_{\mathbf{q}_1}(0) \rangle^{\rm c}_{\rm e}\langle \rho_{\mathbf{q}_2}^{\dagger}(t)\rho_{\mathbf{q}_2}(t) \rangle^{\rm c}_{\rm e}}}$ is the contribution from the opposite momenta term. We assume time-translational symmetry for the system and consequently, the equal-time correlation will be kept constant for $\mathbf{q}_1$ and $\mathbf{q}_2$. We use the product of these equal-time correlations,  $\langle \rho_{\mathbf{q}_1}^{\dagger}(0)\rho_{\mathbf{q}_1}(0) \rangle^{\rm c}_{\rm e}\langle \rho_{\mathbf{q}_2}^{\dagger}(t)\rho_{\mathbf{q}_2}(t) \rangle^{\rm c}_{\rm e}$, as the normalizing factor for the generalized Siegert relation. When restricting the electronic system with spatial translational symmetry, $\braket{ \rho_{\vb{q}_1}^\dagger (t_1) 
    \rho_{\vb{q}_2} (t_2)}_{\rm e}^{\rm c} = 0$ when $\vb{q}_1 \neq \vb{q}_2$,  this generalized form can be reduced back to the original single-momentum Siegert relation in Eq.\,\eqref{eq:SRorigin}.

Without the spatial translational symmetry, the main difference between the current generalized relation and the original Siegert relation comes from the $C^{\rho\rho\rho\rho}_{\mathbf{q}_1\mathbf{q}_2,\rm OM}$ contribution of the autocorrelation spectrum explicitly and leads to the ``classical breakdown'' of the original Siegert relation. This ``classical breakdown'' offers an opportunity to study systems with broken spatial translational symmetry, such as those influenced by external potentials, disorders, or boundary effects.

On the other hand, compared to the classical Siegert relation, the measurement in the quantum regime becomes much more complicated. It involves multiple combinations of the scattering time with the different phase factors. Even if we only compare the quantum contribution from the time-delay component $C^{\rho\rho\rho\rho}_{\mathbf{q}_1\mathbf{q}_2}( t_1, t_2, t_2, t_1)$---which is analogous to the classical response---the charge density operator is non-commutative for the quantum electronic systems. As a result, the contribution of exchanging the operators is quantified by $C^{\rho\rho\rho\rho}_{\mathbf{q}_1\mathbf{q}_2,\rm XC} \neq 0$, which leads to the ``quantum breakdown'' of the original Siegert relation in Eq.\,\eqref{eq:SRorigin} and the generalized relation Eq.\,\eqref{eq:SRtwoMomenta}. This breakdown indicates that XPCS in the quantum regime differs significantly from that in the classical regime. In the following sections, we restrict ourselves to study only the time-delay component to compare with the classical result in non-interacting Fermi gas and the 1D Kitaev chain.

\section{Calculation for the Quantum Electronic System at zero Temperature}\label{sec:qes}

To demonstrate the capabilities of the proposed quantum XPCS, we perform explicit calculations of the four-point density correlation on two prototypical model systems: the non-interacting Fermi gas and the 1D Kitaev chain. The non-interacting Fermi gas serves as a platform to demonstrate how quantum exchange effects can leave distinctive signatures in XPCS, even in the absence of Coulomb correlations. Notably, we show that the Siegert relation breaks down in this simple scenario, while the oscillatory features in the exchange-correlation (XC) channel highlight the promising potential of XPCS to probe exchange-correlation effects. In contrast, to emphasize the power of higher-order correlations detected by quantum XPCS, we apply the method to the 1D Kitaev chain, investigating whether these higher-order signals can reveal signatures of the topological phase.

\begin{figure*}[t]
    \centering   \includegraphics[width=0.95\linewidth]{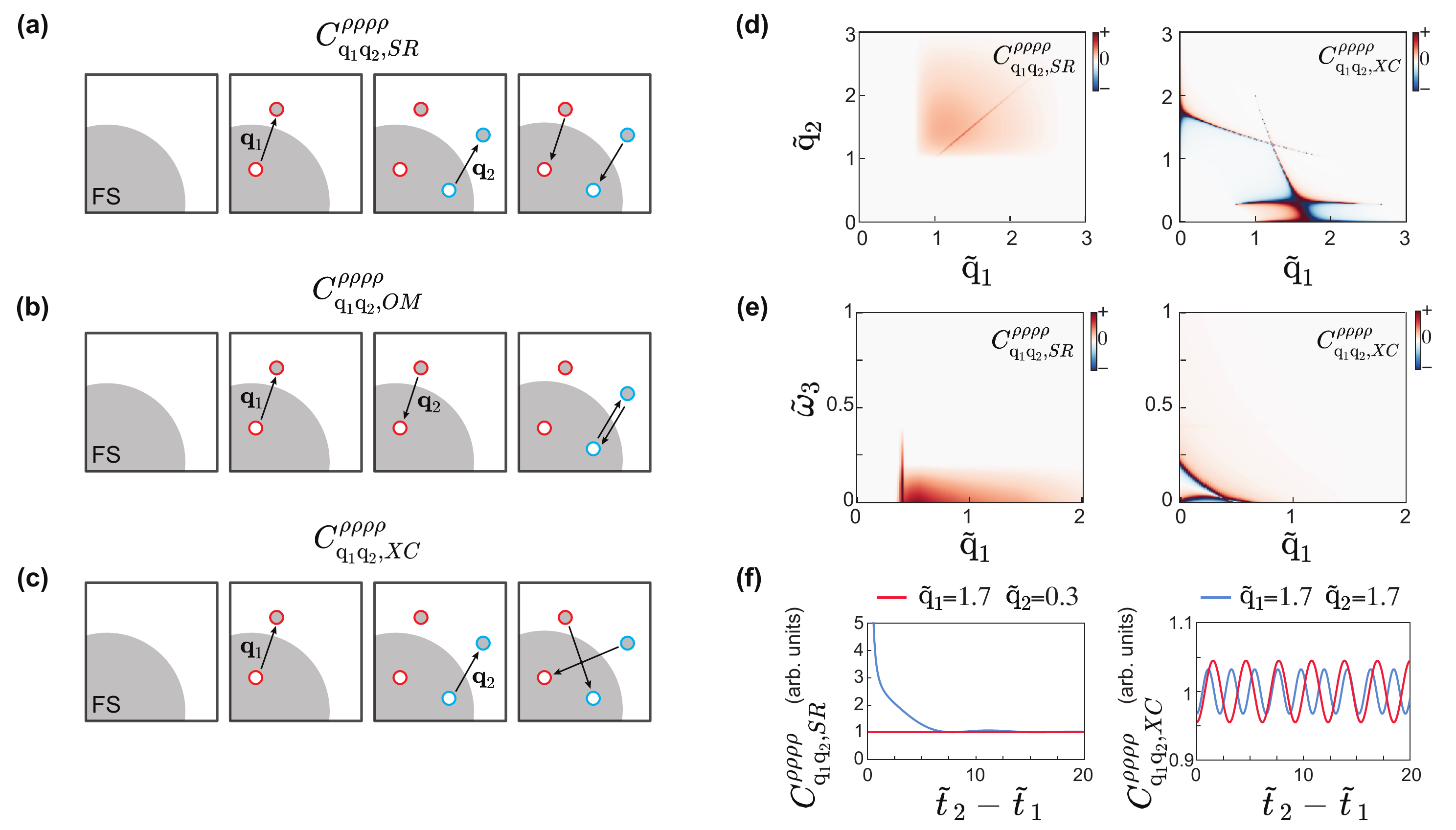}
\caption{Calculated XPCS-accessible time domain four-point density correlation of non-interacting Fermi gas. (a-c) Schematics of the process involved in four-point density correlation of non-interacting Fermi gas contribution, where the gray area represents Fermi sea, and the different colored circles represent the different electron-hole pairs excited by the incoming X-ray photons. $\mathbf{q}_1$ and $\mathbf{q}_2$ with associated 
arrows are the corresponding momentum transfer. (a) $C^{\rho\rho\rho\rho}_{\mathbf{q}_1\mathbf{q}_2, \rm SR}$, where two electron-hole pairs are created independently by two photons then annihilate separately. (b) $C^{\rho\rho\rho\rho}_{\mathbf{q}_1\mathbf{q}_2, \rm OM}$, where one photon contributes to the creation and the other photon contributes the annihilation of one single electron together. (c) $C^{\rho\rho\rho\rho}_{\mathbf{q}_1\mathbf{q}_2, \rm XC}$, where two electron pairs are created independently by two-photon but the first electron (red) annihilate with the second hole (blue), and vice versa. Take $\vb{q}_1 = q_1 \hat{x}$ and $\vb{q}_2 = q_2 \hat{x}$ for simplicity, and normalize the frequency, momentum and time as $\tilde{q}_i = q_i/k_\mathrm{F}$, $\tilde{\omega}_i = \mathrm{\omega}_i/(k_\mathrm{F}^2/ m)$, $\tilde{t}_i  = (k_\mathrm{F}^2/ m) t_i$, where $i = 1, 2$.
    (d-e) $C^{\rho\rho\rho\rho}_{\vb{q}_1\vb{q_2},\rm  SR}$ and $C^{\rho\rho\rho\rho}_{\vb{q}_1\vb{q_2}, \rm XC}$ as a function of (d) $\tilde{q}_1$ and $\tilde{q}_2$ at $\tilde{\omega}_1 = 1.0, \tilde{\omega}_2 = 1.5, \tilde{\omega}_3 = 0$, (e) $\tilde{q}_1$ and $\tilde{\omega}_3$ at $\tilde{\omega}_1 = 0.4 k_\mathrm{F}^2/m$, $\tilde{\omega}_2 = 0.2 k_\mathrm{F}^2/m$, $\tilde{\rm{q}}_2 = 0.4$. (f) $C^{\rho\rho\rho\rho}_{\vb{q}_1\vb{q_2},\rm  SR}(t_1, t_2, t_2, t_1)$ and $C^{\rho\rho\rho\rho}_{\vb{q}_1\vb{q_2},\rm  XC}(t_1, t_2, t_2, t_1)$ as a function of time difference $\tilde{t}_2 - \tilde{t}_1$. $C^{\rho\rho\rho\rho}_{\mathbf{q}_1\mathbf{q}_2,\rm  XC}$ ($C^{\rho\rho\rho\rho}_{\mathbf{q}_1\mathbf{q}_2,\rm  SR}$) are normalized with averaged $C^{\rho\rho\rho\rho}_{\mathbf{q}_1\mathbf{q}_2,\rm  XC}$ ($C^{\rho\rho\rho\rho}_{\mathbf{q}_1\mathbf{q}_2, \rm SR}$) at $t_2 - t_1 \rightarrow \infty$. We note that $C^{\rho\rho\rho\rho}_{\mathbf{q}_1\mathbf{q}_2,\rm  SR}$ is non-zero only in the electron-hole continuum  $\left|q_i - k_\mathrm{F} - \sqrt{2 m \omega_i}\right| < k_\mathrm{F}$. On the other hand, $C^{\rho\rho\rho\rho}_{\mathbf{q}_1\mathbf{q}_2,\rm  XC}$ is non-zero across the momentum domain with the singularity near the resonance when the frequency matches the electron-hole excitation energy. The oscillatory signature offers a means to directly probe electron exchange effect. 
    } 
    \label{fig:FS}
\end{figure*}

\subsection{Exchange-Correlation Contributions in Non-interacting Fermi Gas}

For the non-interacting and spinless Fermi gas, we can consider the ground state wave function $\ket{ \rm FS}_{\rm e} = \prod_{\mathbf{k} \in \rm{FS}} c^\dagger_{\mathbf{k}} \ket{0}_{\rm e}$, which is the ground state of the Hamiltonian,
\begin{align}
    \mathcal{H}_{\rm e}^{\rm FG} = \sum_{\mathbf{k}} c^\dagger_{\mathbf{k}} \epsilon_{\mathbf{k}} c_{\mathbf{k}} ,
\end{align}
where $\epsilon_\mathbf{k} = \frac{|\mathbf{k}|^2}{2m}-\mu$ is the dispersion of free fermion, $\mu$ is the chemical potential, and $m$ is the mass of the fermion. In the non-interacting Fermi gas case, we have the following time-domain Green's functions:
\begin{equation}
    \begin{split}
        iG^{<}_{\mathbf{k}}(t,t') &= -n_\mathrm{F}(\mathbf{k}) e^{-i \epsilon_{\mathbf{k}} (t-t')} ,\\
        iG^{>}_{\mathbf{k}}(t,t') &= (1-n_\mathrm{F}(\mathbf{k})) e^{-i \epsilon_{\mathbf{k}} (t-t')},\\
        G^{T}_{\mathbf{k}}(t,t') &= \Theta(t-t')G^{>}_{\mathbf{k}}(t,t') +\Theta(t'-t)G^{<}_{\mathbf{k}}(t,t') ,\\
        G^{\tilde{T}}_{\mathbf{k}}(t,t') &= \Theta(t'-t)G^{>}_{\mathbf{k}}(t,t') +\Theta(t-t')G^{<}_{\mathbf{k}}(t,t'), \\
    \end{split}
    \end{equation}
where $n_\mathrm{F}$ is the Fermi-Dirac distribution, and the Greens' functions could be Fourier transformed into the frequency domain as:
\begin{equation}
    \begin{split}
        G^{<}_{\mathbf{k}}( \omega)  &=  -2\pi n_\mathrm{F}(\mathbf{k})\delta(\omega - \epsilon_{\mathbf{k}}),\\
        G^{>}_{\mathbf{k}}( \omega) &= -2\pi (1-n_\mathrm{F}(\mathbf{k}))\delta(\omega - \epsilon_{\mathbf{k}}),\\
        G^{T}_{\mathbf{k}}( \omega)  &=   \frac{-n_\mathrm{F}(\mathbf{k})}{\omega - \epsilon_{\mathbf{k}} - i0^{+}} + \frac{1-n_\mathrm{F}(\mathbf{k})}{\omega - \epsilon_{\mathbf{k}} + i0^{+}},\\
        G^{\tilde{T}}_{\mathbf{k}}( \omega) &=  \frac{-n_\mathrm{F}(\mathbf{k})}{\omega - \epsilon_{\mathbf{k}} + i0^{+}} + \frac{1-n_\mathrm{F}(\mathbf{k})}{\omega - \epsilon_{\mathbf{k}} - i0^{+}}.\\
    \end{split}
\end{equation}
The contribution from the fourth-order connected density correlation $C^{\rho\rho\rho\rho}_{\mathbf{q}_1\mathbf{q}_2, \rm XC}$ can be expressed as the sum of six products of Green's functions, that
\begin{align}
    &C^{\rho\rho\rho\rho}_{\mathbf{q}_1\mathbf{q}_2,\rm  XC} (t_1, t_2, t_3, t_4) = \nonumber\\
    &\sum_{\vb{k}}  G^T_{\vb{k} + \vb{q}_1}(t_1, t_2) G^{>}_{\vb{k} + \vb{q}_1 + \vb{q}_2}(t_2, t_3) G^{\tilde{T}}_{\vb{k} + \vb{q}_1}(t_3,  t_4) G^{<}_{\vb{k}} (t_4, t_1)
    \nonumber\\
    &+ G^T_{\vb{k} + \vb{q}_1}(t_1, t_2) G^{>}_{\vb{k + q_1 + q_2}}(t_2, t_4) G^{T}_{\vb{k} + \vb{q}_2}(t_4,  t_3) G^{<}_{\vb{k}} (t_3, t_1)
    \nonumber\\
    &+ G^>_{\vb{k + q_1}}(t_1, t_3) G^{<}_{\vb{k} + \vb{q}_1 - \vb{q}_2} (t_3, t_2) G^{>}_{\vb{k} + \vb{q}_1}(t_2,  t_4) G^{<}_{\vb{k}} (t_4, t_1)
    \nonumber\\
    &+ (\text{time-reversal terms}).
\end{align}

Similarly, the disconnected components can be calculated through Green's functions above \cite{mihaila2011lindhard}.

The disconnected components represent the two-point correlation, which keep the Siegert relation. In contrast, the connected component is the four-point correlation, which represents the correlation that comes from exchanging the quantum operators and can lead to the breakdown of the Siegert relation.
Figs.\,\ref{fig:FS}(a)-\ref{fig:FS}(c) show the schematics of the contribution to the four-point correlation of the density operator, which involves the excitation of two electron-hole pairs and their recombination. The calculated intrinsic four-point density correlation $C^{\rho\rho\rho\rho}_{\mathbf{q}_1\mathbf{q}_2}$ is plotted in Figs.\,\ref{fig:FS}(d)-\ref{fig:FS}(f). 

In the frequency domain, the exchange-correlation channel $C^{\rho\rho\rho\rho}_{\vb{q}_1\vb{q}_2,\rm  XC}$ persists even when the $C^{\rho\rho\rho\rho}_{\vb{q}_1\vb{q}_2,\rm SR}$ term is zero, indicating the power of XPCS to probe quantum exchange-correlation effects. 

In the time domain, $C^{\rho\rho\rho\rho}_{\vb{q}_1\vb{q}_2,\rm  XC}$ shows an oscillatory behavior with a frequency $\vb{q}_1 \cdot \vb{q}_2/m$, while $C^{\rho\rho\rho\rho}_{\vb{q}_1\vb{q}_2,\rm  SR}$ remains constant if $\vb{q}_1 \neq \vb{q}_2$ or decays if $\vb{q}_1 = \vb{q}_2$. The oscillatory nature of the four-point density correlation from the XC channel makes XPCS highly promising for directly probing electron exchange-correlation effects.

\subsection{Detecting Topological Phase with XPCS in 1D Kitaev Chain}
\begin{figure}[t]
    \centering\includegraphics[width=1.00\linewidth]{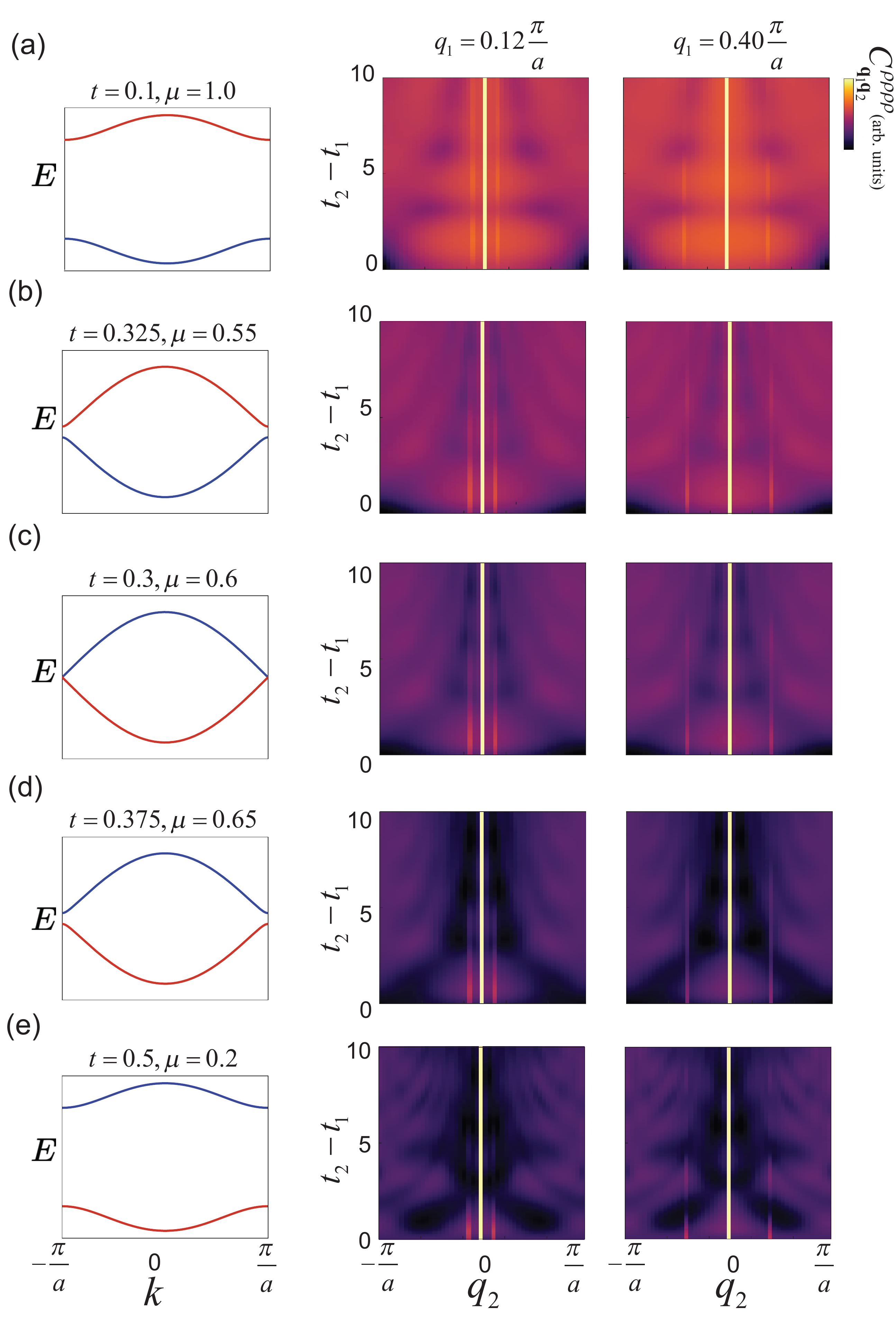}
    \caption{Probing topological phase of a 1D Kitaev chain using XPCS. The band structures (left column) of the 1D Kitaev chain, and the calculated time-domain four-point correlation $C^{\rho\rho\rho\rho}_{\mathbf{q}_1\mathbf{q}_2}(t_1, t_2, t_2, t_1)$ using DMRG, as a function of $t_2 - t_1$ and $q_2$ at two selected $q_1 = 0.12 \frac{\pi}{a}$ (center column) and $q_1 = 0.40 \frac{\pi}{a}$ (right column), where $q_1$ and $q_2$ are the corresponding scalar of $\mathbf{q}_1$ and $\mathbf{q}_2$ in a 1D system. The calculation is performed with the following parameters (a) $t = \Delta = 0.1, \mu = 1.0$, (b) $t = \Delta = 0.275, \mu = 0.65$, (c) $t = \Delta = 0.3, \mu = 0.6$, (d) $t = \Delta = 0.325, \mu = 0.55$, and (e)  $t = \Delta = 0.5, \mu = 0.2$. The color scale is normalized using the elastic signal $q_2 = 0$ and is on a logarithmic scale. DMRG is performed on a $N = 50$ chain.}
    \label{fig:dmrg}
\end{figure}

Next, we consider the 1D Kitaev chain \cite{kitaev2001unpaired}, which serves as a prototypical model for 1D a 
$p$-wave (pw) superconductor. The mean-field Hamiltonian can be written as
\begin{align}
    \mathcal{H}_{\rm e}^{\rm{pw}} = \sum_j  \left( -t c^\dagger_j c_{j+1}  + \Delta  c^\dagger_{j} c^\dagger_{j+1} + h.c. \right) - \mu \sum_{j} c^\dagger_j c_j,
\end{align}
where $t$ is the electron hopping amplitude, $\Delta$ is the superconducting order parameter, $\mu$ is the chemical potential, and $c^\dagger_j(c_j)$ denotes the creation (annihilation) operator of the electron at site $j$. Using the momentum space representation $c_j = \sum_{k} e^{-i {k} a} c_{k}$, where $a$ is the lattice constant, and further defining $\bm{\psi}_k = \begin{pmatrix} \psi_{{ k},+} \\ \psi_{k,-} \end{pmatrix}$, where $\psi_{{ k},+} = u_{ k} c_{ k} + v_{ k} c^\dagger_{ -k}$ and $\psi_{k,-} = -v^*_{ k} c_{ k} + u^*_{ k} c^\dagger_{ -k}$, we can rewrite the Hamiltonian as 
\begin{align}
\mathcal{H}_{\rm e}^{\rm{pw}} & = \sum_{ k} \bm{\psi}_k^\dagger \bm{\epsilon}_{{ k}} \bm{\psi}_k ,
    \\
    \bm{\epsilon}_{{ k,++}} &=\sqrt{(2t\cos ({k}a) + \mu)^2 + (2\Delta\sin({ k}a))^2},\\
    \bm{\epsilon}_{{ k,--}} &=-\sqrt{(2t\cos ({k}a) + \mu)^2 + (2\Delta\sin({ k}a))^2}
\end{align}
Consider the subset of phase space where $\Delta = t$, the chain hosts topological and trivial phases, separated by the band gap closing at $\mu = \pm 2t$. The trivial phase $|\mu| >  2t$ can be adiabatically connected to the atomic insulator where the band is completely filled/empty. On the other hand, the topological phase $|\mu| <  2t$ hosts a zero mode localized at the boundary of the chain, so-called Majorana Zero Mode (MZM) \cite{mandal2023topological}. We have numerically calculated $C^{\rho\rho\rho\rho}_{\mathbf{q}_1\mathbf{q}_2}(t_1, t_2, t_2, t_1)$ using the Density Matrix Renormalization Group (DMRG) method \cite{PhysRevLett.69.2863,RevModPhys.77.259} with the ITensors \cite{itensor} package (Fig.\,\ref{fig:dmrg}) along the transition from topologically trivial phase (Fig.\,\ref{fig:dmrg}(a)) to topological phase (Fig.\,\ref{fig:dmrg}(e)). We note that while the dispersion of the trivial band (Fig.\,\ref{fig:dmrg}(a) left) and topological band (Fig.\,\ref{fig:dmrg}(e) left) can be identical, the correlation functions of each phase are distinct due to the nature of the quasi-particles. For the topologically trivial phase, the excitation can be understood via a localized Wannier function with a tight-binding hopping term. However, as the system enters into the topological phase, the long-range entanglement decreases the single-particle spectral weight, which results in an overall weaker signal. In addition, a momentum-dependent oscillatory pattern emerges in the correlation function in the topological phase, despite the presence of a finite gap in the system. Such phenomenon cannot be attributed to the Friedel oscillation, which rely on the existence of a Fermi surface for the excitation. Instead, the observed oscillatory pattern may reflect unique interference effects of the topological phase in the higher-order correlation function. We speculate that this pattern serves as a distinctive signature of the topological phase in the 1D Kitaev chain, and could potentially be observed in other quantum systems exhibiting topological properties. Such features the promise of the XPCS to probe topological phases beyond existing spectroscopic techniques.

\section{Discussions and Outlook}\label{sec:con}

In this work, we develop a microscopic quantum many-body theory of XPCS that measures the two-photon correlation functions. We propose four XPCS configurations: two ``classical'' configurations based on intensity-intensity correlation and two ``quantum'' configurations inspired by HBT measurements for cleaner higher-order correlation functions. These correlation functions are directly related to the four-point electron density correlations. Even in non-interacting Fermi gas, we show that the measured correlation function breaks the Siegert relation due to Fermionic statistics, manifesting as oscillatory signals on top of the Siegert relation. We also compute the correlation function for the 1D Kitaev chain, revealing qualitative differences in the correlation structure between the topological and trivial phases, even with identical single-particle excitations. These findings support the potential of XPCS as a possible probe for topological phases, including topological orders with long-range entanglement that have long been challenging to identify due to the ambiguous signature in two-point correlation functions \cite{broholm2020quantum}.

 With ever-increasing X-ray coherence and brightness, we propose non-conventional quantum XPCS configurations inspired by HBT from quantum optics to exploit the improved coherence in X-ray sources. Other quantum optics configurations could also inspire experimental setups to characterize additional material properties using X-ray scattering. For instance, if indistinguishable X-ray photons can be generated, Hong-Ou-Mandel configurations \cite{PhysRevLett.59.2044} for photon bunching in higher-order correlation could be realized, while integrating the Mach-Zehnder interferometry \cite{zehnder1891neuer,mach1892ueber} with X-ray momentum dependence could further elucidate electronic interaction mechanisms. In parallel, the improved coherence of cold neutron sources such as Second Target Station may also lead to the probe of other higher-order correlation functions but for spin degrees of freedom. 

Furthermore, the proposed XPCS theory could be generalized to describe techniques such as \textit{in-situ} techniques and pump-probe configurations with additional optical pumps. Previously, \textit{in-situ} techniques have been combined with traditional XPCS to investigate complex polymer dynamics within processing environments \cite{PhysRevMaterials.7.045605} and structural dynamics during additive manufacturing \cite{yavitt2020structural}. The pump-probe technique is extremely powerful for studying phenomena far from equilibrium \cite{grubel2007xpcs,roseker2011development}, particularly light-induced phenomena such as Floquet-engineered topological phases \cite{PhysRevB.79.081406,PhysRevB.84.235108,bukov2015universal,mciver2020light} and light-induced superconductivity \cite{fausti2011light,mitrano2016possible}. We anticipate that the quantum XPCS in conjunction with a pump-probe configuration will facilitate the characterization of electron dynamics with unprecedented details.

However, while XPCS may have the potential to elucidate electron dynamics, processing high-volume and high-dimensional XPCS data presents significant challenges that call for enhanced data interpretation. Recently, supervised learning methods have been applied to identify topological phases of Majorana zero modes from tunneling spectroscopy signals \cite{cheng2024machine}, and an unsupervised deep learning framework has been developed to automatically classify relaxation dynamics in XPCS data \cite{horwath2024ai,andrejevic2023data}. We expect the development of efficient machine learning methods to play an indispensable role in XPCS data interpretation and analysis. With rapid advancements in future X-ray and neutron facilities and high-performance computation for data processing, this theoretical foundation can establish a cornerstone for future higher-order correlation measurements of quantum materials, leveraging the advanced capabilities of X-ray and neutron facilities effectively.

\section*{Acknowledgement}
The authors thank C Brohlom, M Chan, L Chapon, X Chen, Z Chen, J Dilling, E Fradkin, M Greven, J Moore, J Taylor, A Tennant, J Turner, Y Wang, and H Zhou for the insightful discussions. PS acknowledges support from the US Department of Energy
(DOE), Office of Science (SC), Basic Energy Sciences (BES), Award No. DE-SC0020148. CF acknowledges support from DOE Award No. DE-SC0021940. M Landry acknowledges support from the National Science Foundation
(NSF) Convergence Accelerator Award No. 2235945. RO and DCC acknowledge the support from Designing Materials to Revolutionize and Engineer our
Future (DMREF) Program with Award No. DMR-2118448. YW is supported by the U.S. DOE BES, under Early Career Award No.~DE-SC0024524. M Li is partially supported by the Class of 1947 Career Development Chair and support from R Wachnik. 

\appendix

\section{Intensity operator and detector resolution}
To describe the detector signal in terms of the intensity operator, we can consider the intensity readout as the number of photons measured in a given time frame. The average intensity can be described as an operator 
\begin{align}
    \bar{I}_{\vb{q}_1 \bm{\eps}_1}(t) = \int dt_1 I_{\vb{q}_1\bm{\eps}_1}(t_1) \tilde{\delta}(t - t_1),
\end{align}
where $\tilde{\delta}(t)$ describe the time window of photon measurement. The ideal detector limit reduces the function to the Dirac-delta function. For a detector that detects photons in the time window $T_w$, we can write $\hat{\delta}(t) = \Theta(T_w/2 - |t|)/T_w$. The detector measurement reflects the eigenvalue of $\bar{I}(t)$, and the autocorrelation is described as the convolution of the intensity correlation with the detector resolution,
\begin{align}
    &\langle \bar{I}_{\vb{q}_1\bm{\eps}_1}(t) \bar{I}_{\vb{q}_2\bm{\eps}_2}(t') \rangle = \nonumber 
    \\ & \int dt_1 dt_2 \langle I_{\vb{q}_1\bm{\eps}_1}(t_1) I_{\vb{q}_2\bm{\eps}_2}(t_2) \rangle \tilde{\delta}(t - t_1)\tilde{\delta}(t - t_2).
\end{align}
With this formulation, we can see that the dynamics on a timescale much faster than the window $T_w$ will get averaged out and can be approximated by the ensemble average. Hence, the measured intensity reflects only the slow time-scale dynamics where the fast dynamics are averaged out. 

Alternatively, we can consider the inelastic X-ray scattering setup where the photon count is accumulated over a long time scale which can be calculated via the Fermi Golden rule,
\begin{align}
    \braket{I_{\vb{q}\bm{\eps}_1}} = 2 \pi | \langle \Psi_f |  \mathcal{H}_{\rm{int}} | \Psi_i \rangle |^2 \delta((\omega_{\vb{p}_{\rm{in}}} - \omega_\vb{p}) - (E_f - E_i)).
\end{align}

\section{Details of the two-point correlation function calculation}
\label{Supp:single-photon}

Rewrite the light-matter interaction explicitly as
\begin{gather}
     \sH_{\rm int} = \sum_{\bm{\eps}_1\bm{\eps}_2 \mathbf{p}_1 \bm{p}_2} \hat\sD_{\mathbf{p}_1\mathbf{p}_2}^{\bm{\eps}_1\bm{\varepsilon}_2} a^\dagger_{\mathbf{p}_2\bm{\eps}_2}a_{\mathbf{p}_1\bm{\eps}_1},\\ \hat\sD_{\mathbf{p}_1 \mathbf{p}_2}^{\bm{\eps}_1\bm{\eps}_2} = \sum_{\alpha\beta \mathbf{k}} \sM^{\alpha\beta\bm{\varepsilon}_1\bm{\varepsilon}_2}_{\mathbf{kp}_1\mathbf{p}_2} c^\dagger_{\mathbf{k}+\mathbf{p}_1-\mathbf{p}_2,\beta}c_{\mathbf{k},\alpha},
\end{gather}
where we assume resonances can be neglected. 
To simplify the following derivation, we work in the interaction picture and define 
\be
     \tilde{\sH}_{\rm int}(t) = e^{i\sH_0 t/ \hbar} \sH_{\rm int} e^{-i\sH_0t/ \hbar}. 
\ee
 
Then the time-evolution operator in the interaction picture takes the form
\be
     U(t,t_0) = \hat 1 - {\frac{i}{\hbar}} \int_{t_0}^t dt_1 \tilde{\sH}_{\rm int}(t_1)+\cdots , 
\ee
and the intensity of scattered light with momentum $\vb{p}_1$~is
\begin{align}
	I_{\mathbf{q}_1\bm{\eps}_1}(t) =   a^\dagger_{\mathbf{p}_1\bm{\eps}_1}(t) a_{\mathbf{p}_1\bm{\eps}_1}(t),
\end{align}
where $\vb{p}_{\rm in}$ is the momentum of incoming light and $\vb{q}_1=\vb{p}_{\rm in} - \vb{p}_1$ is the momentum transfer. 
The initial state is $\ket\Psi\equiv \ket\phi_{\rm e}\otimes \ket \phi_{\rm ph}$ defined at time $t_0$ (typically we take $t_0\to-\infty$) in the interaction picture, where $\ket\phi_{\rm e}$ is the electronic ground state and $\ket\phi_{\rm ph}\equiv \bar{D}(s) \ket 0$ is the photon coherent state, with $\bar{D}(s) = \exp \Bigl( \sum_{\mathbf{p}\bm{\eps}} \bigl(s_{\mathbf{p}\bm{\eps}}a^\dagger_{\mathbf{p}\bm{\eps}}-s^{*}_{\mathbf{p}\bm{\eps}}a_{\mathbf{p}\bm{\eps}}\bigr) \Bigr)$ as the displacement operator. Then the single photon intensity measurement can be expanded as

\begin{gather}
     \vev{I_{\vb{q}_1} (t)} = \bra{\Psi} U^\dagger(t,t_0) a_{\vb{p}_1\bm{\eps}_1 }^\dagger a_{\vb{p}_1\bm{\eps}_1 }U(t,t_0) \ket{\Psi}  \\
     = \vev{\bar{D}(s) U^\dagger(t,t_0)a^\dagger_{\vb{p}_1\bm{\eps}_1} a_{\vb{p}_1\bm{\eps}_1} U(t,t_0) \bar{D}(s)}  \\ 
     \begin{split}
    \approx &  \frac{1}{\hbar^2} \int_{t_0}^t dt_1 \int_{t_0}^t dt_2  
      \Big \langle {\bar{D}(s) \tilde  \sH_{\rm int} (t_1)
      a^\dagger_{\vb{p}_1\bm{\eps}_1} 
     a_{\vb{p}_1\bm{\eps}_1} 
     \tilde\sH_{\rm int}(t_2)  \bar{D}(s)} \Big\rangle
     \end{split}\\
     \begin{split}
     = &  \frac{1}{\hbar^2} \int_{t_0}^t dt_1 \int_{t_0}^t dt_2 e^{i\omega_{\mathbf{p}_1}(t_2-t_1)}  \sum_{ \bm{\varepsilon}_{i1,2} \bm{\varepsilon}_{f1,2}' \atop\mathbf{k}_{1,2} \mathbf{k}_{1,2}'}\vev{\hat\sD_{\mathbf{k}_1\mathbf{k}_1'}^{\bm{\varepsilon}_{i1}\bm{\varepsilon}_{f1}}(t_1) \hat\sD_{\mathbf{k}_{f2}\mathbf{k}_{i2}}^{\bm{\varepsilon}_{f2} \bm{\varepsilon}_{i2}}(t_2)} \\
     & \times  \vev{\bar{D}(s^{t_1})  a^\dagger_{\mathbf{k}_1\bm{\varepsilon}_{i1}} a_{\mathbf{k}_1'\bm{\varepsilon}_{f1}} a^\dagger_{\vb{p}_1\bm{\eps}_1} a_{\mathbf{p}_1\bm{\eps}_1} a_{\mathbf{k}_2'\bm{\varepsilon}_{f2}}^\dagger a_{\mathbf{k}_2\bm{\eps}_{i2}}  \bar{D}(s^{t_2})},
     \end{split}
\end{gather}
where $\sH_0$ is the free Hamiltonian, $s^t_{\mathbf{q}\bm{\eps}}\equiv e^{-i\omega_{\mathbf{q}} t} s_{\mathbf{q}\bm{\eps}}$, and we use the time-evolution identity for coherent photon states $e^{-i\sH_0t/\hbar} \bar{D}(s) e^{i \sH_0 t /\hbar} = \bar{D}(s^t)$. In the third equality above, we drop all terms that don't involve a factor of the interaction Hamiltonian (this is justified by noting that the detector is very far away and does not detect the initial coherent state).

Now, continuing to suppose that our detector is very far away, the only way for photons to reach it is by scattering off of the target electrons.  As a result, the only terms of the above that contribute come from contracting creation/annihilation operators in the intensity operator $I_{\mathbf{q}_1}$ with those of the interaction Hamiltonian $\sH_{\rm int}$. We therefore have
\begin{gather}
      \vev{\bar{D}(s^{t_1})  a^\dagger_{\mathbf{k}_1\bm{\varepsilon}_{i1}} a_{\mathbf{k}_1'\bm{\varepsilon}_{f1}} a^\dagger_{\vb{p}_1\bm{\eps}_1} a_{\mathbf{p}_1\bm{\eps}_1} a_{\mathbf{k}_2'\bm{\varepsilon}_{f2}}^\dagger a_{\mathbf{k}_2\bm{\eps}_{i2}}  \bar{D}(s^{t_2})}\nonumber \\
       \rightarrow \vev {\bar{D}(s^{t_1}) a^\dagger_{\mathbf{k}_1\bm{\varepsilon}_{i1}} a_{\mathbf{k}_2\bm{\varepsilon}_{i2}} \bar{D}(s^{t_2})} \delta_{\bm{\varepsilon}_{f1} \bm{\eps}_1} \delta_{\bm{\varepsilon}_{f2} \bm{\eps}_1} \delta_{\mathbf{k}_1' \mathbf{p}_1}\delta_{\mathbf{k}_2' \mathbf{p}_1 } \nonumber \\
       = \bar s_{\mathbf{k}_1\bm{\varepsilon}_{i1}}^{t_1} s_{\mathbf{k}_2\bm{\varepsilon}_{i2}}^{t_2} \delta_{\bm{\varepsilon}_{f1} \bm{\eps}_1} \delta_{\bm{\varepsilon}_{f2} \bm{\eps}_1} \delta_{\mathbf{k}_1' \mathbf{p}_1}\delta_{\mathbf{k}_2' \mathbf{p}_1 },
\end{gather}
which yields 
\begin{gather}
\begin{split}
     \vev{I_{\vb{q}_1\bm{\eps}_1}(t) } =  \frac{1}{\hbar^2}\int_{t_0}^t dt_1 \int_{t_0}^t dt_2  \sum_{\mathbf{k}_{1,2}}\vev{\hat\sD_{\mathbf{k}_1 \mathbf{p}_1}^{\bm{\eps}_{\rm in }\bm{\varepsilon}_1} (t_1) \hat\sD_{\mathbf{p}_1 \mathbf{k}_2}^{\bm{\eps}_{2} \bm{\eps}_{\rm in} }(t_2)} \\ 
     \times e^{i\omega_{\vb{p}_1}(t_2-t_1)} \bar s_{\mathbf{k}_1\bm{\varepsilon}_{\rm in}}  s_{\mathbf{k}_2\bm{\varepsilon}_{\rm in}} e^{i\omega_{\mathbf{k}_1} t_1} e^{-i\omega_{\mathbf{k}_2} t_2} .
\end{split}
\end{gather}
Noting that $s_{\vb{k}\bm{\eps}}$ only depends on $\vb{k}^\parallel = (\omega_{\vb{k}} /c)  \hat{\vb{p}}_{\rm in}$---that is, the component of $\vb{k}$ parallel to $\vb{p}_{\rm in}$---we may define $\tilde s_{\bm{\eps}}(\omega_{\vb k}) \equiv s_{\vb{k}\bm{\eps}}$. Supposing that the incoming photons have polarization $\bm{\eps}$, the above then reduces to
\begin{gather}\begin{split}
     \vev{I_{\vb{q}_1\bm{\eps}_1} (t)} = \frac{1}{\hbar^2}\int_{t_0}^t dt_1 \int_{t_0}^t dt_2  \sum_{\vb{k}_1^\parallel \vb{k}_2^\parallel}  \vev{\hat\sD_{\mathbf{k}_1 \mathbf{p}_1}^{\bm{\eps}_{\rm in }\bm{\varepsilon}_1} (t_1) \hat\sD_{\mathbf{p}_1 \mathbf{k}_2}^{\bm{\eps}_{2} \bm{\varepsilon}_{\rm in} }(t_2)} \\\times e^{i\omega_{\vb{p}_1}(t_2-t_1)}\tilde s_{\bm{\varepsilon}_{\rm in }}(-\omega_{\vb{k}_1}) \tilde s_{\bm{\varepsilon}_{\rm in}}(\omega_{\vb{k}_2})  e^{i\omega_{\vb{k}_1} t_1} e^{-i\omega_{\vb{k}_2} t_2} ,
\end{split}\end{gather}
where $\vb{k}_{1,2}^\parallel$ are the components of $\vb{k}_{1,2}$ parallel to $\vb{p}_{\rm in}.$
Now let us use the fact that $\tilde s_{\bm{\varepsilon}}(\omega)$ is peaked very sharply at $\omega=\omega_{\rm{in}}$: assuming $\mathbf{q}_1$ is much larger than the momentum space width of $\tilde s$, we may replace $\vb{k}_{1,2}^\parallel\to \mathbf{p}_{\rm{in}}$ in the $\hat \sD$'s. As a result, the sum over $\vb{k}_{1,2}^\parallel$ gives an effective Fourier transform on $\tilde s$ 
\begin{gather}
     \sum_{\vb{k}_\parallel} \tilde s_{\bm{\varepsilon}_{\rm in}}(\omega_{\vb{k}_\parallel}) e^{-i\omega_{\vb{k}_\parallel} t} \to  {V \ov \sigma_{\rm beam}} \int{d \omega\ov 2\pi c } \tilde s_{\bm{\varepsilon}_{\rm in}}(\omega_{\vb{k}_\parallel}) e^{-i\omega_{\vb{k}_\parallel} t} \cr= s_{\bm{\varepsilon}_{\rm in}}(t) e^{-i\omega_{\rm in} t}.
\end{gather}
We thus have
\begin{gather}\begin{split}
     \vev{I_{\vb{q}_1\bm{\varepsilon}_{1}}(t)  }=  \frac{1}{\hbar^2}\int_{t_0}^t dt_1 \int_{t_0}^t dt_2  \vev{\hat\sD_{\mathbf{k}_1 \mathbf{p}_1}^{\bm{\eps}_{\rm in }\bm{\varepsilon}_1} (t_1) \hat\sD_{\mathbf{p}_1 \mathbf{k}_2}^{\bm{\eps}_{2} \bm{\varepsilon}_{\rm in} }(t_2)} \\ \times  s_{\bm{\varepsilon}_{\rm in}}(t_1)  s_{\bm{\varepsilon}_{\rm in}}(t_2)  e^{i(\omega_{\rm{in}}-\omega_{\vb{p}_1}) (t_1-t_2)}  .
\end{split}\end{gather}
Lastly, notice that transnational symmetry ensures that the $\hat\sD$'s are functions of $\vb{p}-\vb{p}_{\rm in} = \vb{q}_1$ only. Neglecting polarization effects, we may then define $\sD_{\vb{q}_1} \equiv \hat\sD_{\vb{p}_1 \vb{p}_{\rm in}}$, and making the approximation~\eqref{sDapprox}, we recover~\eqref{single}. Finally, in deriving this expression we began by supposing the electrons started in their ground state. To account for a finite-temperature system simply take the expectation value above $\vev{\cdots}$ with respect to a thermal electronic state.

\section{DMRG calculation for four-point correlation of 1D Kitaev chain}
\begin{figure}
\centering\includegraphics[width=1.0\linewidth]{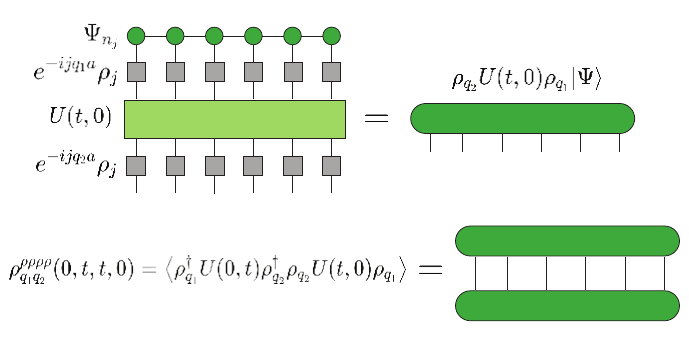}
    \caption{The schematics of $C^{\rho\rho\rho\rho}_{\text{q}_1\text{q}_2}$ calculation from the original MPS state with $\rho_{\text{q}_1}$, $\rho_{\text{q}_2}$, and the time evolution.}
    \label{fig:MPS}
\end{figure}

The Density Matrix Renormalization Group (DMRG) is a highly efficient variational optimization algorithm for determining the ground states of Hamiltonians in low-dimensional quantum many-body systems, achieving unprecedented precision, particularly for one-dimensional quantum systems \cite{PhysRevLett.69.2863,RevModPhys.77.259}. This method generates the optimized matrix product state (MPS), which is recognized as an efficient representation of the ground states of gapped local Hamiltonians \cite{fannes1992finitely}. To search for the ground state of the 1D Kitaev chain using DMRG, the state is represented as an MPS:

\begin{align}
    \Psi_{n_1...n_N} = \langle n_1 ... n_N | \Psi \rangle  = \sum_{\{s_i: i \in [N]\}} \psi_{n_1}^{(s_1)}\psi_{n2}^{(s_1 s_2)}  ... \psi_{n_N}^{(s_{N-1})},
\end{align}
 where $n_i$ is the occupation at site $i$, and $s_i$ is the related site indices. The ground state is calculated using the variational principles to minimize $\braket{\mathcal{H}_{\rm e}}_{\rm e}$. With the MPS representation, the time-evolution of this state can be computed under the dynamics of a Hamiltonian through the ``time-evolving block decimation'' (TEBD) \cite{PhysRevLett.93.040502}. To calculate the four-point correlations $C^{\rho\rho\rho\rho}_{\mathbf{q}_1\mathbf{q}_2}(0, t, t, 0)$, we rewrite and observe the correlation with the form of the norm of the wavefunction
\begin{align}
    C^{\rho\rho\rho\rho}_{\mathbf{q}_1\mathbf{q}_2}(0, t, t, 0) &= \langle \rho^\dagger_{\mathbf{q}_1}U(0, t)\rho_{\mathbf{q}_2}^\dagger\rho_{\mathbf{q}_2}U(t, 0)\rho_{\mathbf{q}_1}\rangle 
    &= \langle \tilde{\Psi} | \tilde{\Psi} \rangle.
\end{align}
Here $\ket{\tilde{\Psi}} = \rho_{\mathbf{q}_2}U(t, 0)\rho_{\mathbf{q}_1}|\Psi\rangle$ can be represented as the photon interacted MPS state with the time evolution in Fig.\,\ref{fig:MPS}, and the time domain XPCS signal for 1D Kitaev superconducting chain can be calculated.

\bibliography{refs_paper}

\end{document}